# Unlocking the Structural Mystery of Vaterite CaCO$_3$

Xingyuan San[1†], Junwei Hu[2†], Mingyi Chen[2], Haiyang Niu[2*], Paul J. M. Smeets[3], Jie Deng[4], Kunmo Koo[3], Roberto dos Reis[3], Vinayak P. Dravid[3*], Xiaobing Hu[3*]

[1]Hebei Key Lab of Optic-electronic Information and Materials, The College of Physics Science and Technology, Hebei University, Baoding 071002, China

[2]State Key Laboratory of Solidification Processing, International Center for Materials Discovery, School of Materials Science and Engineering, Northwestern Polytechnical University, Xi'an 710072, China

[3]Department of Materials Science and Engineering, The NU*ANCE* Center, Northwestern University, Evanston, IL 60208, USA

[4]Department of Geosciences, Princeton University, Princeton, NJ 08544, USA

[†]These authors contributed equally to this work.

[*]Corresponding authors. Email: xbhu@northwestern.edu (X.H.); haiyang.niu@nwpu.edu.cn (H.N.); v-dravid@northwestern.edu (V.P.D.)

**Abstract**

Calcium carbonate (CaCO$_3$), the most abundant biogenic mineral on earth, plays a crucial role in various fields. Of the four polymorphs, calcite, aragonite, vaterite, and amorphous CaCO$_3$, vaterite is the most enigmatic one due to an ongoing debate regarding its structure that has persisted for nearly a century. In this work, based on systematic transmission electron microscopy characterizations, elaborate crystallographic analysis and machine learning aided molecular dynamics simulations with *ab initio* accuracy, we reveal that vaterite can be regarded as a polytypic structure. The basic phase is a monoclinic lattice possessing pseudohexagonal symmetry. Direct imaging and atomic-scale simulations provide evidence that a single grain of vaterite can have three orientation variants. Additionally, we find that vaterite undergoes a second-order phase transition. These atomic scale insights provide a comprehensive understanding of the structure of vaterite and offer new perspectives on the biomineralization process of calcium carbonate.

**Introduction**

Calcium carbonate ($CaCO_3$) is one of the most abundant natural materials found in biomineral systems, which plays a pivotal role in the chemistry of the hydrosphere, lithosphere, biosphere, climate regulation, and scale formation (*1-4*). Among the three anhydrous crystalline structures, vaterite (*5*), aragonite (*6*) and calcite (*7*), vaterite has the most confined stability field. It is frequently observed as a transient intermediate phase during the transition process of amorphous calcium carbonate to other more stable polymorphs (*8, 9*) or formed *via* liquid–liquid phase separation in supersaturated $CaCO_3$ solution (*10-12*), thereby kinetically directing the mineralization pathway. Although vaterite is less stable and rarer compared to other forms of $CaCO_3$ due to its relatively low stability, it can still be commonly found in many biogenic and abiotic systems, such as freshwater lackluster pearls (*13*), cement (*14*) and natural minerals (*15*). In addition, vaterite's unique structure, high solubility in water, and porosity properties relative to other crystalline polymorphs make it an ideal functional material for environmental chemistry (*16, 17*), bone tissue (*18, 19*) and biomedical engineering (*5, 20*). Therefore, an accurate description of the structure of vaterite at the atomic scale is very demanding for gaining the rationale behind the growth dynamics of calcium carbonate (*10, 21-23*) and fulfilling its potential applications as a functional material (*24-26*).

Despite decades of experimental and theoretical efforts (*27-31*), the precise structure of vaterite remains a subject of debate (Supplementary Note I, Table S1). Several conflicting structures have been proposed, such as the hexagonal Meyer lattice (*32*), orthorhombic lattice (*33, 34*), monoclinic lattice and triclinic lattice (*29, 30*). The discrepancies arise not only from the ordering of carbonate groups, particularly the slight rotation within a single layer, but also from the long-range stacking sequence of carbonate layers along the *z*-axis concerning a hexagonal lattice. Recently, by means of aberration-corrected high-resolution transmission electron microscopy (HRTEM), Pokroy *et al* revealed that vaterite crystals contain two interspersed crystal structures, namely a predominant hexagonal Kamhi lattice together with an unknown structure. Although a combined Kamhi and Meyer model can be used to explain the strong diffraction spots, neither one can explain any of the weak diffractions (*31*). Thus, the real structural features of vaterite are still a mystery to humankind. Until now, it has been generally acknowledged that vaterite cannot be described using a single lattice, and there are likely many planar faults including micro twinning and other stacking orders and/or disorders (*27, 33, 35*). However, to our knowledge, direct imaging evidence of various stacking disorders at the atomic scale in vaterite is still missing.

Uncovering the structural features of vaterite through experimental means is an extremely challenging task. Conventional bulk measurements such as X-ray diffraction (XRD), electron diffraction and Raman spectra are not sufficient, and atomic resolution imaging is a prerequisite. HRTEM imaging is one choice, however, this technique is actually very limited in resolving the complex structure here since it can only provide a phase contrast image that usually contains many artifacts caused by thickness and focus variations. In contrast, atomic resolution high angle annular dark field (HAADF) imaging is an ideal technique since it provides the atomic number (*Z*) contrast with the intensity in proportion to $Z^{1.7-1.9}$ (*36*). Here, we uncover the structural mystery of vaterite by combining atomic resolution experiments based on samples from two different sources, the lackluster pearls and the asteriscus of carp, and molecular dynamics simulations adopting a deep neural network (DNN) interatomic potential with *ab initio* accuracy. The results of the two paradigms are complementary and can be cross validated. Our atomic scale insights into the vaterite structure resolve the contentions that existed among previous research work and provide an intrinsic perspective for studying the biomineralization mechanism of $CaCO_3$, which may consequently facilitate the synthesis of calcium carbonate based materials with target structure and performance.

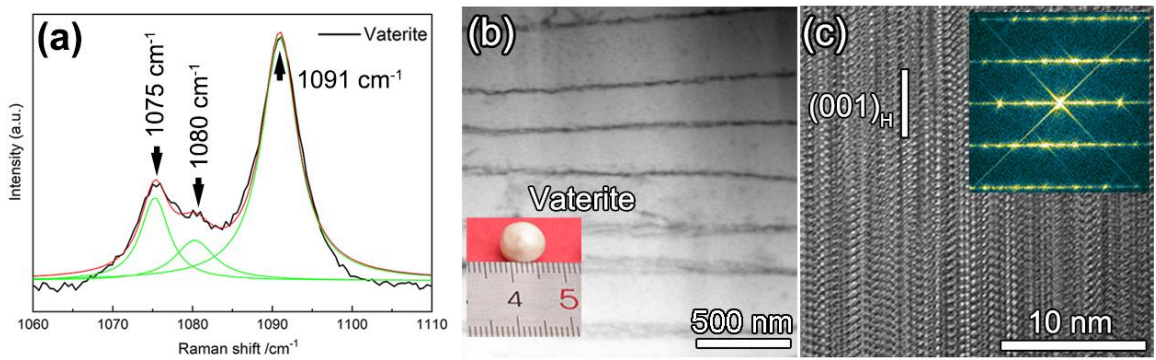

**Fig. 1. Microstructural features of vaterite.** (a) Splitting of the most intense Raman band $v_1$. The best result of the Gaussian fitting procedure was achieved for the decomposition of three peaks at 1075 cm$^{-1}$, 1080 cm$^{-1}$ and 1091 cm$^{-1}$. (b) HAADF image showing the laminate structure. The inset shows the morphological features of the peals with vaterite structure. (c) HRTEM image and inset FFT patterns showing the representative faulted features of vaterite.

**Results and discussion**

**General structural features of vaterite**

We first measured the Raman spectra of the lackluster pearls. As shown in Fig. 1a and Extended Data Fig. 1a, the splitting Raman band $v_1$, which represents the symmetric stretching mode of the carbonate ions, has the highest intensity. The spectral features shown here are a good match to those obtained from geological, synthetic and other biogenic samples comprising vaterite minerals (*37-39*), indicating that our lackluster sample can be categorized as a common biomineral containing a vaterite structure. The most intense peak is at 1091 cm$^{-1}$, with consecutively lower intensities for the peaks at 1075 cm$^{-1}$ and 1080 cm$^{-1}$. The triple splitting suggests the presence of at least three crystallographically independent carbonate groups and similar carbonate group layers within vaterite (*40, 41*). In addition, as shown in Extended Data Fig. 1b, our XRD data agree well with other diffraction data of vaterite in the literature (*37-39*), indicating again that our lackluster pearls are representative biominerals containing a vaterite structure.

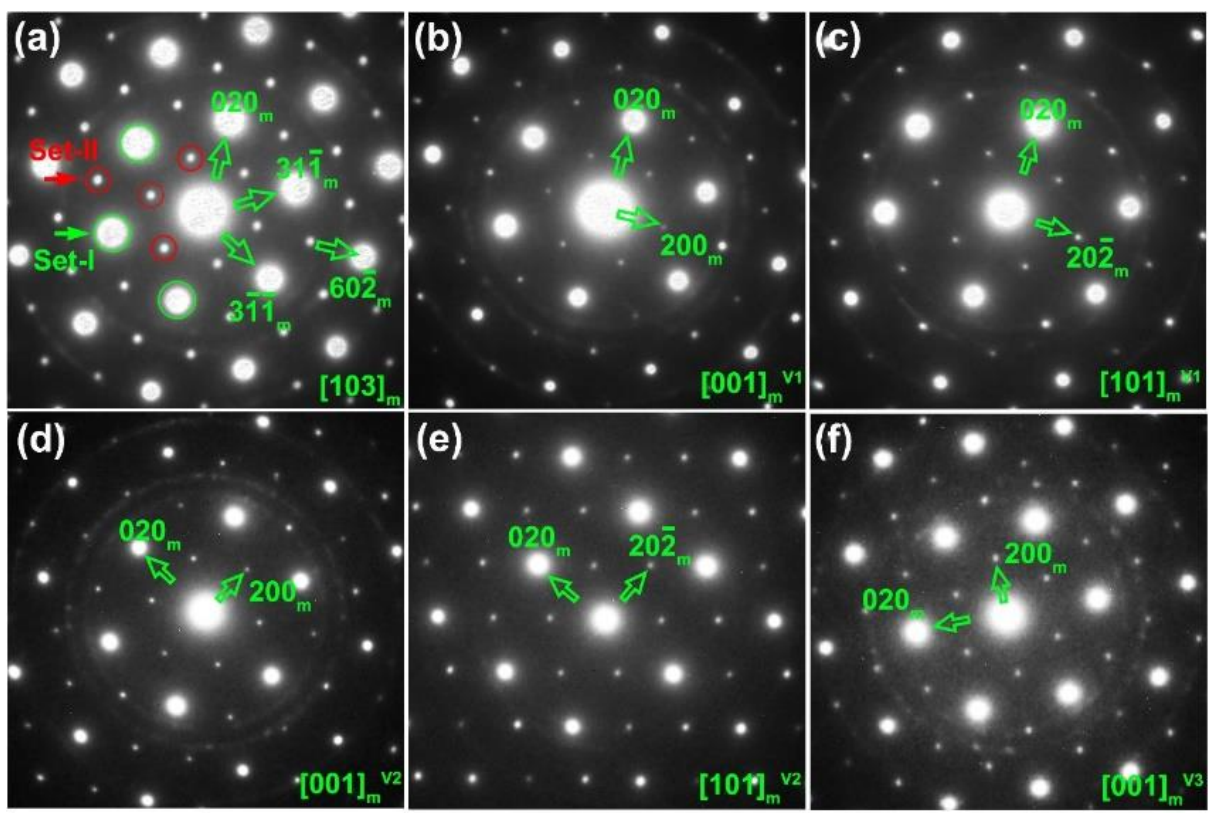

**Fig. 2. A series of SAED patterns of vaterite obtained from a single grain**. EDPs along the (a) $[103]_m$, (b) $[001]_m^{V1}$, (c) $[101]_m^{V1}$, (d) $[001]_m^{V2}$, (e) $[101]_m^{V2}$ and (f) $[001]_m^{V3}$ directions. The subscript *m* represents the monoclinic lattice. The superscripts *V1*, *V2* and *V3* indicate three orientational variants. Circles in (a) indicate two sets of reflections with significant differences in intensities with Set I/II patterns having higher/lower intensities.

Figure 1b presents a high-angle annular dark field (HAADF) image of the lackluster pearl, which shows a laminar structure at the micrometer scale. The associated elemental maps, shown in Extended Data Fig. 1c-e, confirm the uniform distribution of C, Ca and O in the grain interior. Additionally, the HRTEM image in Fig. 1c reveals the general stacking feature of vaterite at the atomic scale. The inset digital fast Fourier transformed (FFT) patterns suggest the presence of many disordered stacking layers along the $(001)_H$ plane, labeled based on the hexagonal Meyer lattice (*32*). This finding is consistent with previous results on other vaterite samples from different originations (*30, 31, 35*). However, due to the significant offset resulting from the phase contrast, the HRTEM image in Fig. 1c cannot be used to extract further structural details.

To uncover the intrinsic structural features of vaterite, we then performed large angle tilting experiments to obtain serial selected area electron diffraction (SAED) patterns from a single grain. The relative positions and associated experimental tilting angles of each diffraction pattern in reciprocal space are given schematically in Supplementary Fig. S1, and diffraction patterns are displayed in Fig. 2. The diffraction patterns in Fig. 2a reveal a hexagonal symmetry feature, which is further confirmed by convergent beam electron diffraction and atomic resolution imaging analysis (Extended Data Fig. 2). Such results agree well with the suggested the hexagonal structure (*42, 43*). However, the diffraction patterns in Fig. 2b-f can only be indexed by a monoclinic structure with lattice parameters of *a*=12.2 Å, *b*=7.2 Å, *c*= 9.3 Å and *β*=115.2°. These lattice parameters could match any of three monoclinic lattices including the *C2*, *Cc*, and *C2/c* structures proposed in previous work (*29*). Unfortunately, diffraction data (Supplementary Fig. S4, S5 and Note III) cannot distinguish between the different rotations of carbonates of these three structures. The structural projection of the *C2* structure along the [010] direction is illustrated in Extended Data Fig. 3a, which can also be described using a hexagonal lattice as highlighted by the dashed line. In fact, the aforementioned monoclinic lattice is closely related to the disordered Meyer structure with a hexagonal lattice (P6$_3$/mmc; *a=b=7.169* Å, *c=16.98* Å). Their intrinsic relationships are discussed in Supplementary Note IV. As shown in Extended Data Fig. 3b, calcium ions ($Ca^{2+}$) approximately form a hexagonal array, and carbonate ions ($CO_3^{2-}$) occupy half of the Ca$_6$ trigonal prismatic interstices in one layer and the other half in the neighboring layers. In each interstice, the carbonate has three possible orientations, and the carbonate ions within a plane are regularly distributed resulting in an ordered arrangement that is more energetically favored (*44*). Considering the rotational invariant, the stacking sequence composed of two adjacent layers is always identical (Supplementary Fig. S10). Once a third layer is introduced, three possible stacking sequences can be obtained, *i.e.*, the carbonate groups in the third layer with three possible orientations correspond to rotations of +120°, -120°, or 0° around the *z*-axis compared to the counterparts in the first layer (Supplementary Fig. S11). The three stacking sequences can be marked as "+", "−", and "0", respectively, following the notation originally proposed by Christy (*28*) (Supplementary Note II and Extended Data Fig. 3c). The *C2* phase shown in Extended Data Fig. 3a and its two analogous structures, namely the *Cc* and *C2/c* structures share the same stacking sequence (+−+−+−), only the rotation of carbonates is slightly different.

In Fig. 2a, the electron diffraction patterns (EDPs) show two sets of reflections with different intensities, as indicated by circles. Only the spots with brighter contrast (Set-I patterns) match the monoclinic structure along the [103]$_m$ direction. In contrast, all of the reflections in Fig. 2b-f can be solely indexed along either the [001]$_m$ or [101]$_m$ direction based on the monoclinic lattice, despite the obvious intensity difference. Importantly, the relative distributions of the reflections in Fig. 2b, 2d and 2f are the same, except for a 60° rotation along the viewing direction. This phenomenon is also observed in Fig. 2c and 2e, indicating that there are likely three orientation variants of the monoclinic lattice within the vaterite structure (Supplementary Fig. S6 and Note V). These three variants can be formed by a 60° rotation along the [103]$_m$ zone-axis, considering the pseudohexagonal features of the monoclinic lattice. The stacking of variant segments along the [103]$_m$ direction yields the stacking disorder observed in Fig. 1c and the weaker spots in Fig. 2a-2f (Supplementary Note VI). Therefore, the domain concept proposed here can well rationalize the broadly existed weak spots that could not be explained in previous work (*31*). Direct evidence of orientation variants in real space will be discussed below.

**Localized structural intergrowth within vaterite**

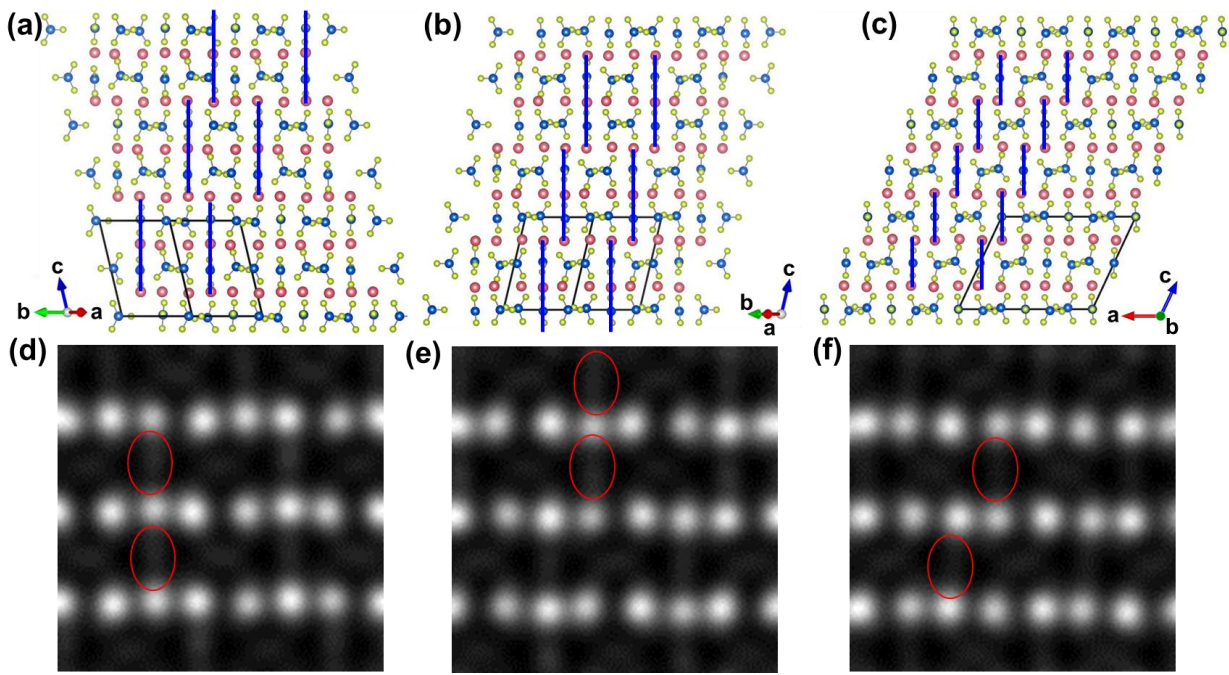

**Fig. 3. Similarity and nuance difference for the stacking of carbonates along some directions.** Structural projection of the ordered *C2* vaterite along the (a) $[110]_m$, (b) $[1\bar{1}0]_m$, and (c) $[010]_m$ directions. The black frames indicate the projection of a unit cell along the corresponding directions. The blue vertical lines indicate the stacking feature of the Ca-C-O chains along the $[001]_m$ direction. Atomic resolution HAADF image simulations along the (d) $[110]_m$, (e) $[1\bar{1}0]_m$, and (f) $[010]_m$ directions. The red circles in (d-f) indicate the diffused streaks along the vertical direction.

To visualize the potential stacking order and/or disorder within vaterite, it is necessary to tilt the grain to the $[010]_m$ zone-axis, as this direction allows for edged-on projection of the planar defects. To facilitate experimental data analysis, we would like to uncover the slight difference along some $[010]_m$ related directions first. As demonstrated in Supplementary Fig. S6 schematic, the monoclinic lattice has a pseudohexagonal feature with a pseudo-6-fold axis along the $[103]_m$ direction. The $[010]_m$ direction is equivalent to the $[110]_m$ and $[1\bar{1}0]_m$ directions. Simulated EDPs along the above three directions are shown in Supplementary Fig. S9a-c, and structural projections of the above three directions are shown in Fig. 3a-c. It is observed that the stacking features of the Ca layers are nearly identical along the above three directions. In-between two neighboring Ca layers along the vertical direction, there are layers composed of carbonates. However, along these projected directions, the stacking features of the carbonate differ (indicated by vertical lines).

Based on the atomic resolution HAADF image simulations shown in Fig. 3d-f, it is found that the C-O bonds orientated approximately along the vertical directions can generate the diffused weak streak features indicated by red circles. Therefore, the stacking features of the diffused streaks can be used to differentiate the stacking features of the carbonate layers.

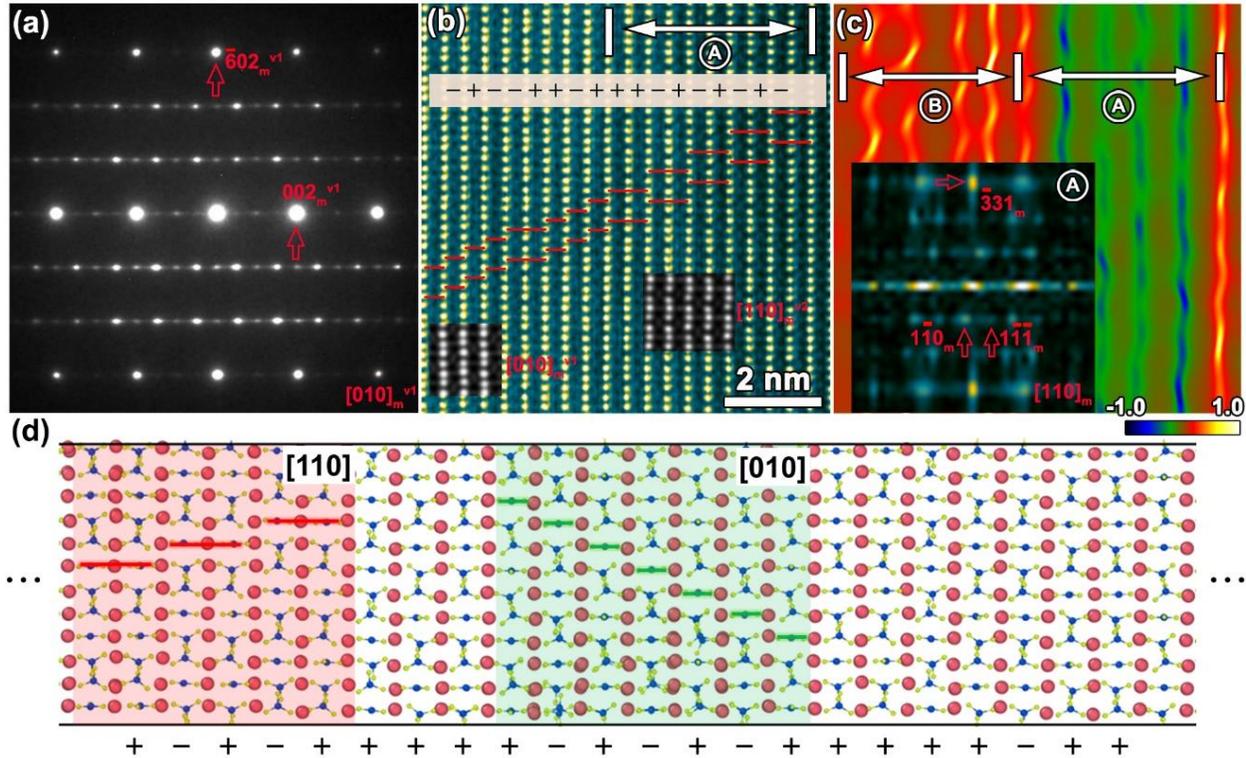

**Fig. 4. Stacking disorder within vaterite.** (a) EDPs obtained from vaterite along the $[010]_m$ zone-axis. (b) Atomic resolution HAADF image along the $[010]_m$ zone axis showing the polytypic features within vaterite. The red vertical lines indicate the stacking feature of the Ca-C-O chains along the $[001]_m$ direction. Inset image simulations correspond to the vaterite structure with the *C2* space group along the $[110]_m$ and $[010]_m$ directions. (c) Strain map ($\varepsilon_{xy}$) of the image shown in (b) obtained by geometric phase analysis (GPA). Inset FFT patterns corresponding to areas A. (d) Theoretically grown vaterite with a polytypic structure. The symbols ("+" and "−") inserted in (b) and (d) indicate the stacking sequences of carbonate layers introduced in Extended Data Fig. 3.

In Fig. 4a, the experimental EDPs of vaterite are shown along the $[010]_m$ zone-axis, indicating a significant amount of stacking disorder on the $(001)_m$ plane due to the strong streaking features observed on this plane. Figure 4b displays an atomic resolution HAADF image along this direction, while the corresponding strain map derived from the geometrical phase analysis (GPA) is shown in Fig. 4c. The interface separating structures of different lattices are apparent, with the right and left regions labeled as A and B, respectively. The inset FFT patterns suggest that A can be indexed solely based on the monoclinic lattice along the $[010]_m$ direction. Within the vaterite grain interior, there are three variants, with variants II and III corresponding to the $[110]_m$ and $[1\bar{1}0]_m$ directions, respectively, if variant I is tilted to the $[010]_m$ direction. While structural projections along these three directions are very similar, as shown in Fig. 3, slightly different stacking sequences of carbonate layers, located between the two calcium layers, can be used to differentiate the variations in the local structural segment. Figure 4b shows that some regions correspond to the monoclinic structure projected along the $[010]_m$ direction, while others correspond to the structural projection of the monoclinic lattice along the $[110]_m$ direction. The atomic resolution HAADF image with a larger field of view, shown in Supplementary Fig. S12, clearly demonstrates that the vaterite structure should be regarded as a polytypic structure with three orientation variants formed based on the monoclinic lattice. These variant segments, together with the stacking disorder, are interrelated along the $[103]_m$ direction, as shown schematically in Extended Data Fig. 4.

It is worth noting that the atomic scale structural features obtained from lackluster pearls are reproducible in other vaterite samples of different originations. Specifically, we have compared characterization results with another vaterite biomineral, an asteriscus present in the otolith of carp, for further analysis. As demonstrated in Supplementary Data Fig. 5, the Raman spectrum of the asteriscus shows typical features of vaterite, and the HRTEM images and associated FFT patterns demonstrate the heavily faulted feature of vaterite grains, similar to what we found in the lackluster sample. Additionally, the atomic resolution HAADF image in Extended Fig. 5d shows the localized stacking features of calcium and carbonate layers, which can also be labelled based on the signs introduced in Extended Data Fig. 3. This direct evidence confirms that the primary structural information obtained from the lackluster pearl is the intrinsic information of vaterite and is generally applicable to other minerals containing a vaterite structure.

To validate our experimental results, we conducted molecular dynamics (MD) simulations to study the crystallization process of vaterite directly from liquid. We trained a DNN potential under the framework of the DeePMD approach (*45*), which can simulate a large system with *ab initio* accuracy for a long time (*46-48*), which is essential for crystallization simulation. To overcome the nucleation barrier, solid–liquid coexistence simulations (Extended Data Fig. 6) were performed, starting with a single fixed Ca solid layer at 1000 $K$. The crystallization proceeded with the reorientation of the carbonate groups, and once completed, the system was quenched to 10 $K$ to eliminate thermal fluctuations. One of the final configurations is shown in Fig. 4d. Our simulation results reveal that the stacking sequence "0" is absent, while "+" and "−" are randomly distributed along the $z$-axis, equivalent to the $[103]_m$ direction in the monoclinic lattice (normal direction of the $(001)_m$ plane). These results are in excellent agreement with our experimental findings shown in Fig. 4b, Supplementary Fig. S12 and Extended Data Fig. 5. Additionally, we obtained structural segments with the same features as the structural projection of the monoclinic lattices along the $[010]_m$ and $[110]_m$ directions, supporting our conclusion from Fig. 4b. We also estimated the total energy of different stacking structures at the same temperature to understand the absence of the stacking sequence "0". The results indicate that the stacking sequence containing "0" significantly increases the potential energy of the system by 41 $meV$ per $CaCO_3$ unit compared to those containing only "+" and/or "−" (Supplementary Note VII and Fig. S14). Therefore, any structure with the stacking sequence of "0" is energetically unfavorable. The EDPs simulations based on the theoretically grown structure shown in Fig. 4d agree well with the experimental observations along the stacking direction, namely, the $[103]_m$ zone axis (Supplementary Fig. S15).

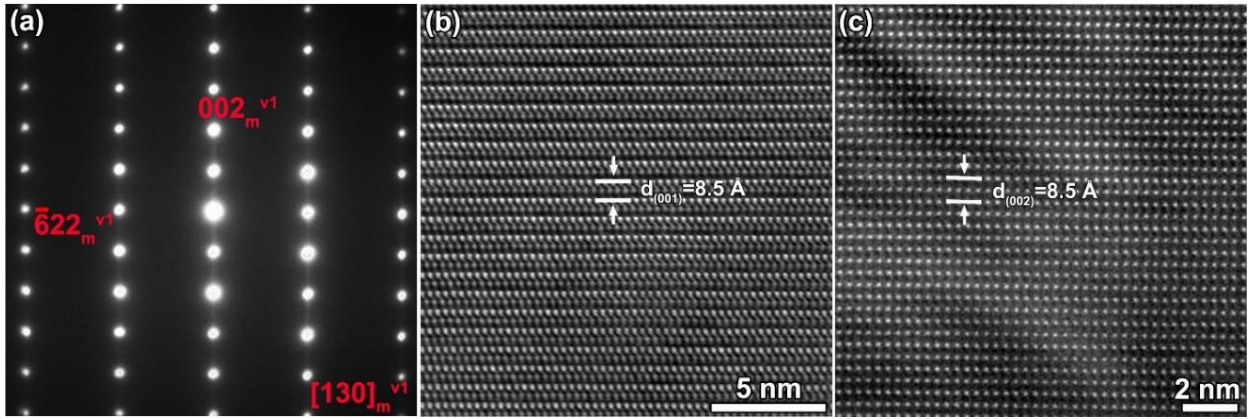

**Fig. 5. Structural projection of vaterite along $[130]_m$ direction.** (a) SAED patterns. (b) Atomic resolution TEM image. (c) Atomic resolution HAADF image.

After tilting the grain by an additional 30° along the $(001)_m^*$ direction, the resulting $[130]_m$ zone-axis is obtained. As depicted in Fig. 5a, the SAED patterns indicate that the weak spots and diffused streaking feature between $(001)_m$ and $(002)_m$, which were observed in EDPs along the [010]$_m$ direction (Fig. 4a), are almost absent. Additionally, the atomic resolution TEM (Fig. 5b) and HAADF images (Fig. 5c) do not exhibit any evident stacking disorders. This can also be rationalized by the concept of orientation variants, where tilting variant I to the [130]$_m$ direction results in variants II and III being tilted to [100]$_m$ and [1$\bar{3}$0]$_m$, respectively. The simulated EDPs along these three orientations are identical (Supplementary Fig. S16). By overlapping the EDPs along these three orientations, the experimental observations shown in Fig. 5a can be reproduced.

**Second-order phase transition in vaterite**

After clarifying the stacking principles of carbonate layers along the *z*-axis (Supplementary Note VII and Fig. S14), the question remains as to how the carbonate groups align in a single layer. Previous vaterite structure models (*29, 34*), including the *C2* structure, show that the carbonate groups tilt slightly away from their high-symmetry orientations. Thus, to unveil the real orientations of the carbonate groups in vaterite, we conducted a detailed analysis of their atomic positions, using the *C2* phase as the initial structure. We observed a small rotation of carbonate as the temperature increases, with the *C2* phase transforming into the *C2/c* phase at approximately 190 *K*, resulting in a higher symmetry. To quantify this phenomenon, we introduced an order parameter, *Q*, which measures the difference in the coordinates of the oxygen atoms in adjacent carbonate groups (see methods and Supplementary Note VIII). As shown in Fig. 6a, by periodically increasing/decreasing the temperature of the system, we find that *Q* fluctuates around zero at high temperature, corresponding to a high-symmetry (HS) state (i.e., *C2/c*), and deviates away from zero at low temperatures, corresponding to a low-symmetry (LS) state (i.e., *C2*). Figure 6b illustrates the orientations of carbonate groups in the HS and LS states. It is worth noting that, the other low-symmetry structure *Cc* was not observed in our simulations, as it would transform into the *C2* structure at a relatively low temperature due to its lower stability.

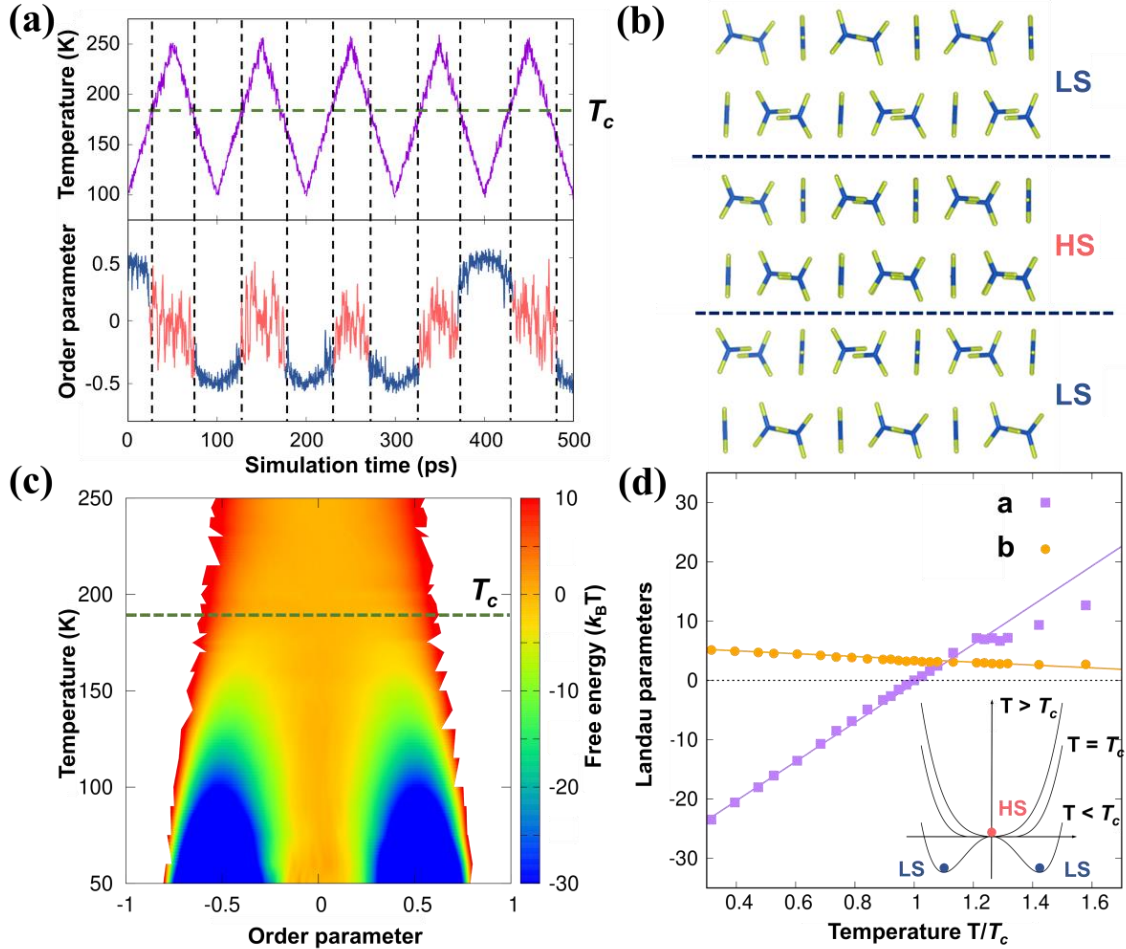

**Fig. 6. Second-order phase transition in vaterite.** (a) Temperature and order parameter $Q$ as a function of simulation time. (b) Schematic illustration of the carbonate group orientations in high-symmetry (HS) and low-symmetry (LS) states. (c) Two-dimensional free energy surface of vaterite as functions of order parameter $Q$ and temperature. (d) Landau parameters as a function of temperature. Inset: Ginzburg-Landau free energy surface below, at, and above the critical temperature $T_c$ of vaterite.

To understand the kinetic correlation between the two states, the well-tempered metadynamics (*49*), a smoothly converging and tunable free-energy method, was employed with the order parameter $Q$ serving as the collective variable. Figure 6c plots the free energy surface as a function of $Q$ and temperature. At high temperatures, a parabola-like free energy surface is observed, while at low temperatures, an energy barrier separates two local minima with the opposite order parameter value. This energy barrier decreases continuously with increasing temperature and eventually disappears at the critical temperature $T_c$ (190 ± 10 K), consistent with the findings presented in Fig. 6a. At the same time, the order parameter decreases continuously to zero, indicating a second-order phase transition in the Ginzburg-Landau model (*50*).

We find the free energy can be well described by $F(T,Q) = aQ^2 + bQ^4$, where $a$ and $b$ are phenomenological quantities known as Landau parameters. As predicted by the Ginzburg-Landau model, $a$ exhibits a linear temperature dependence near the critical point $T_c$, i.e., $a(T) = \alpha(T - T_c)$, while parameter $b$ is a constant. We estimated the Landau parameters by fitting the free energy surface in Fig. 6c and the results are shown in Fig. 6d. When the temperature is around $T_c$, both parameters can be well fitted. At relatively high temperatures, $a(T)$ starts to exhibit nonlinearity. Although $b(T)$ is not a perfect constant, the slope of its dependence on temperature is relatively small, and such characteristic has been reported in previous studies (*51, 52*). At $T_c$, the low energy area of the free energy surface becomes rather flat. We find that the LS structures with positive and negative Q values are mirror-symmetrized to each other and share the same space group *C2*. As demonstrated in Fig. 6a, the LS structures with different chirality can interconvert between each other passing through the HS structure. Such chirality feature in vaterite might be significant in some specific biosystems, in which one enantiomer might be favored (*53*).

**Conclusions**

In summary, a comprehensive investigation of biogenic vaterite from two different marine species, the lackluster pearls and the asteriscus of carp, was conducted by combining systematic TEM characterizations, crystallographic analyses, and advanced theoretical calculations. Eventually, the structural mystery of vaterite CaCO$_3$ is uncovered. The basic structure of vaterite is a monoclinic lattice with a space group of *C2* or *C2/c*, with lattice parameters of *a*=12.3 Å, *b*=7.13 Å, *c*=9.4 Å, and *β*=115.480°. The difference between the *C2* and *C2/c* structures is characterized by the slightly different orientations of the carbonate groups within a single layer, which are sensitive to temperature, leading to different symmetries in vaterite. Importantly, the transformation between the low- and high-symmetry phases can be well described by a second-order phase transition with a critical point of approximately 190 *K*. Vaterite has a pseudohexagonal feature with a pseudo sixfold axis along the [103]$_m$ direction, and three variants that are formed due to a 60° rotation along this direction. In principle, vaterite is a polytypical structure formed by the nanoscale random intergrowth of the three orientation variants, which are interrelated along the [103]$_m$ direction. The stacking sequence "0" is absent, while "+" and "−" sequences are randomly distributed along the $[103]_m$ direction based on the monoclinic lattice. Additionally, the localized structural and chemical features of minerals are correlated with environmental nucleation and growth conditions, such as temperature, pressure and presence of impurities. However, the intrinsic structural features of vaterite are similar and the configuration rules do not change. The atomic scale insights of vaterite obtained from lackluster pearls and the asteriscus in the otoliths of carp are generally applicable to all the vaterite samples from different origination.

**Materials and Methods**

**Sample preparation and electron microscopy methods**

The lackluster pearl (vaterite structure) and normal pearl (aragonite structure) were collected from Zhuji, Southeast China. The asteriscus otolith pairs were extracted from the carps originated at Baiyang lake in China. The sample was cut into different sections using a linear precision saw with a thickness of ~500 *μm* and then ground to ~100 *μm* using silicon carbide (SiC) papers. Thin foils with a diameter of < 3 *mm* were stuck on the molybdenum grid, then grounded using variant grit SiC papers, and polished with diamond paste to ~10 *μm* and finally thinned by Ar ion milling in a Gatan precision ion polishing system (PIPS) using a low voltage (3-4 *kV*) to avoid possible ion beam damage. Furthermore, to improve the electronic conductivity, perforated samples were sputtered with carbon layer using a sputter machine (Quorum Q 150R ES) for ~20 *s*.

Raman spectra were obtained in back-scattering mode at room temperature by utilizing the Jobin-Yvon HR800 micro-Raman system with a grating speed of 1800 *lines/mm*. This system was equipped with a liquid-nitrogen-cooled charge-coupled device (CCD) and a ×100 objective lens (NA=0.9). The resolution of the Raman system was 0.35 $cm^{-1}$ per pixel. Raman spectra were measured using a 532 nm laser. A tilt series of SAED patterns were recorded using a FEI Tecnai T12 BioTWIN at 120 *kV* with a $LaB_6$ thermionic gun. The tilting angles along the X/Y axis range from -70°/-35° to +70°/+35°. Atomic scale resolution TEM and HAADF images were acquired using a JEM ARM200CF microscope which was operated at 200 *kV*. This microscope was equipped with a cold field-emission gun, a probe corrector, and the dual silicon drift detectors (SDDs). The point resolution in scanning TEM (STEM) mode is ~78 *pm*. The convergence semi-angle for STEM imaging was ~22 *mrad*. The inner and outer acceptance semi-angles for HAADF imaging were approximately 90 and 250 *mrad*, respectively. Within this collection angle range, the intensity of images is dominated by incoherent *Z* contrast. The area of a single detector is 100 $mm^2$ and the solid angle of this energy dispersive spectrum (EDS) system is approximately 1.8 steradian (*sr*).

Simulations of the EDPs are performed using the CrystalDiffract software. Atomic resolution HAADF image simulations were performed using the Dr. Probe software (*54*). The simulation parameters were chosen close to the experimental conditions assuming that all aberrations have been corrected within the probe forming aperture corresponding to a 21 *mrad* semi-convergence angle. The annular detector collection angles range from 80 to 200 *mrad*. An effective Gaussian source profile of 0.1 *nm* full width at half maximum (FWHM) was utilized. Simulated images were obtained as series over sample thicknesses up to 10 *nm*.

**Theoretical calculations**

*Ab initio molecular dynamics calculation*

The *ab initio* molecular dynamics (AIMD) calculations were performed using the Vienna *ab initio* package (VASP) (*55, 56*). The general gradient approximation of Perdew-Burke-Ernzerhof (GGA-PBE) was adopted for the exchange-correlation functional (*57, 58*). The electron wave function was expanded using a plane wave with a cut-off energy of 400 *eV*. Reciprocal space integration was performed by setting the *k* spacing to 0.5 (*59*). The AIMD was carried out with an isothermal-isobaric (NPT) ensemble at the temperature of 1000 *K* using the Langevin thermostat (*60*) and at the pressure of one bar using the Parrinello-Rahman barostat (*61, 62*). The time step in the simulation was 1 *fs*.

*Density functional theory (DFT) energy and force calculations*

Single point energy and forces were calculated using VASP. The smearing parameter and exchange-correlation density functional were the same as those in the AIMD simulation, while the energy cut-off for this calculation was increased to 1000 *eV*.

*Training deep neural network model*

All training of calcium carbonate DNN models was performed with the package DeePMD-kit (*45*). The settings of the training parameters are as follows: the cut-off radius smoothly decays from 7.5 $Å$ to 8 $Å$. The sizes of three hidden layers for the embedding network and three hidden layers for the fitting network were set to (20, 40, 80) and (240, 240, 240) respectively. The learning rate decays from $1.0 \times 10^{-3}$ to $5.3 \times 10^{-8}$. The pre-factors of the energy, force and virial term in the loss function change from 0.04 to 0.2, 1000 to 1 and 0.04 to 0.2, respectively. The total number of training steps was set to $1 \times 10^6$.

To establish a training set capable of covering the whole solidification process, first, we collected configurations from the AIMD trajectories, in which we fixed half of the calcium atoms in the supercell while carbonate ions were allowed to rotate freely at extremely high temperatures to obtain various liquid–solid interface structures. Approximately 13,000 configurations were obtained together with the corresponding energy, force, and virial data as the training set to establish the first-generation DNN model (DNN1). We next performed MD simulations on Lammps (*63*) using DNN1 to enable adequate sampling of the configuration space at larger timescales. To make our DNN model accurately predict the energies and forces of the structures under various thermodynamic conditions, the MD simulations covered the temperature range 1-2000 *K*. Additionally, initial solid structures with different stacking sequences along the *z*-axis were also considered. Notably, the orientations of carbonates varied casually at high temperature before melting, which indicated that not only the structures with ordered carbonates distributed over three orientations in one layer, but also the structures with disordered carbonates were included in our training set.

Based on the configurations obtained in the previous steps, we performed DFT single point energy calculations and constructed a new training set to establish DNN2. We repeated the above steps to perform MD simulations under the corresponding thermodynamic conditions and calculated the single point energies for the configurations extracted from the trajectories. The energies and forces calculated by DFT were compared with the predicted values of DNN (Supplementary Fig. S13), and the configurations with large error were added to the training set for iterative optimization. The errors of the final DNN model in terms of the atomic energies and forces on the test sets were $1.81 \times 10^{-3}$ *eV/atom* and *0.11 eV/Å* respectively.

*Molecular dynamics simulations*

The molecular dynamics simulations were carried out using Lammps, integrated with DeePMD-kit to utilize the DNN potential. The simulation system with 6480 atoms (1296 $CaCO_3$ units), consisting of 36 layers of calcium atoms and 36 layers of carbonates, was in an orthorhombic box with periodic boundary conditions in all three directions with the dimensions of $2.2017 \times 2.5406 \times 15.3321$ *nm³*.

To perform the following crystallization simulations, we first constructed a liquid–solid coexisting configuration in which the solid phase was composed of only one single layer of vaterite. This was done through a simulation at a high temperature of 1600 *K* using the NPT ensemble starting from a perfect structure of vaterite, in which one layer of calcium atoms was fixed. Thus, the ordered calcium carbonates transformed into disordered states except for the fixed layer of calcium atoms. Since calcium atoms exhibit a hexagonal crystal structure in vaterite, fixing a layer of calcium atoms would allow the system to grow more easily in the specific direction. Then, we used the obtained configuration to perform the crystallization simulations. The liquid–solid phase transition proceeded along the [001] direction of the hexagonal lattice during 1000 *K* relaxation of the NPT mode. Our structure arises from the natural transformation kinetics and therefore indicates the intrinsic properties of the structure of vaterite alone the [001] direction. After obtaining a complete solid phase, we further quench the system to 10 $K$ in 1 $ns$ to reduce the thermal fluctuations. Snapshots of the trajectory are presented in Extended Data Fig. 6. Several simulations under the same thermodynamic conditions, but the velocities with different random seeds, were performed to cross-validate the simulation results.

In the above simulations, temperature and pressure were controlled by the stochastic velocity rescaling thermostat (*64*) and the Parrinello-Rahman barostat, respectively. The timestep is set to 1 *fs*. The initial velocity of each atom was generated with the Boltzmann distribution. The linear momentum was rescaled by subtracting the momentum of the center of the mass every 1000 timesteps. Plumed (*65*) packaged in Lammps was used for performing well-tempered metadynamics (*66*) and analyzing the trajectories. Crystal structure visualization was achieved using Ovito (*67*) and Vesta (*68*).

*Well-tempered metadynamics and free energy surface*

To estimate the difference in free energy between the high symmetry (HS) phase and the low symmetry (LS) phase at different temperatures, we further performed well-tempered metadynamics (WTMetaD) with the order parameter $Q$ introduced in Supplementary Note VIII. The bias potential of WTMetaD is a history-dependent function of the order parameter composed by the deposits of Gaussians, and the height of the additive Gaussian dwindles as the sampling proceeds. For the comprehensive descriptions please refer to (*68*). In this work, the width and the initial height of Gaussian are set to 0.075 $Å$ and 3 *kJ/mol*, respectively. The bias factor, which characterizes the rate of change of the deposited Gaussian height, is set to 30. The time interval of the deposits of Gaussians is equal to 500 *fs*. We calculated the free energy surfaces at 24 different temperatures with a temperature interval of 10 $K$ from 50 $K$ to 250 $K$. The other MD settings for the simulations were the same as those used in the growth process as described above. The simulation time was set as $2\times10^6$ steps (2 *ns*) to reach convergence. After the sampling was finished, we reweighted the obtained data using an on-the-fly strategy (*69*) to evaluate the free energy surfaces.

**References**

1. R. E. Zeebe, J. C. Zachos, K. Caldeira, T. Tyrrell, Carbon emissions and acidification. *Science* **321**, 51-52 (2008).
2. J. MacAdam, S. A. Parsons, Calcium carbonate scale formation and control. *Rev. Environ. Sci. Biotechnol.* **3**, 159-169 (2004).
3. F. C. Meldrum, Calcium carbonate in biomineralisation and biomimetic chemistry. *Int. Mater. Rev.* **48**, 187-224 (2003).
4. Z. Zou *et al.*, A hydrated crystalline calcium carbonate phase: Calcium carbonate hemihydrate. *Science* **363**, 396-400 (2019).
5. B. V. Parakhonskiy, A. Haase, R. Antolini, Sub-micrometer vaterite containers: synthesis, substance loading, and release. *Angew. Chem. Int. Ed. Engl.* **51**, 1195-1197 (2012).
6. X. San *et al.*, Uncovering the crystal defects within aragonite $CaCO_3$. *Proc. Natl. Acad. Sci. U.S.A.* **119**, e2122218119 (2022).
7. Y. Y. Kim *et al.*, Tuning hardness in calcite by incorporation of amino acids. *Nat. Mater.* **15**, 903-910 (2016).
8. J. D. Rodriguez-Blanco, S. Shaw, L. G. Benning, The kinetics and mechanisms of amorphous calcium carbonate (ACC) crystallization to calcite, viavaterite. *Nanoscale* **3**, 265-271 (2011).
9. E. Seknazi, S. Mijowska, I. Polishchuk, B. Pokroy, Incorporation of organic and inorganic impurities into the lattice of metastable vaterite. *Inorg. Chem. Front.* **6**, 2696-2703 (2019).
10. P. J. M. Smeets, K. R. Cho, R. G. Kempen, N. A. Sommerdijk, J. J. De Yoreo, Calcium carbonate nucleation driven by ion binding in a biomimetic matrix revealed by *in situ* electron microscopy. *Nat. Mater.* **14**, 394-399 (2015).
11. P. J. M. Smeets *et al.*, A classical view on nonclassical nucleation. *Proc. Natl. Acad. Sci. U.S.A.* **114**, E7882-E7890 (2017).
12. A. F. Wallace *et al.*, Microscopic evidence for liquid-liquid separation in supersaturated $CaCO_3$ solutions. *Science* **341**, 885-889 (2013).
13. L. Qiao, Q.-L. Feng, Z. Li, Special vaterite found in freshwater lackluster pearls. *Cryst. Growth. Des.* **7**, 275-279 (2007).
14. T. Saito, E. Sakai, M. Morioka, N. Otsuki, Carbonation of γ-$Ca_2SiO_4$ and the mechanism of vaterite formation. *J. Adv. Concr. Technol.* **8**, 273-280 (2010).
15. J. D. C. McConnell, Vaterite from Ballycraigy, Larne, Northern Ireland. *Miner. Mag. J. Miner. Soc.* **32**, 535-544 (1960).
16. X. Song, Y. Cao, X. Bu, X. Luo, Porous vaterite and cubic calcite aggregated calcium carbonate obtained from steamed ammonia liquid waste for $Cu^{2+}$ heavy metal ions removal by adsorption process. *Appl. Surf. Sci.* **536**, (2021).
17. K. G. M. D. Abeykoon, S. P. Dunuweera, D. N. D. Liyanage, R. M. G. Rajapakse, Removal of fluoride from aqueous solution by porous vaterite calcium carbonate nanoparticles. *Mater. Res. Express* **7**, 035009 (2020).
18. A. Obata, H. Ozasa, T. Kasuga, J. R. Jones, Cotton wool-like poly(lactic acid)/vaterite composite scaffolds releasing soluble silica for bone tissue engineering. *J. Mater. Sci. Mater. Med.* **24**, 1649-1658 (2013).
19. Q. Huang, Y. Liu, Z. Ouyang, Q. Feng, Comparing the regeneration potential between PLLA/Aragonite and PLLA/Vaterite pearl composite scaffolds in rabbit radius segmental bone defects. *Bioact. Mater.* **5**, 980-989 (2020).
20. D. B. Trushina, T. V. Bukreeva, M. V. Kovalchuk, M. N. Antipina, $CaCO_3$ vaterite microparticles for biomedical and personal care applications. *Mater. Sci. Eng. C Mater. Biol. Appl.* **45**, 644-658 (2014).

21. D. Gebauer, A. Völkel, H. Cölfen, Stable prenucleation calcium carbonate clusters. *Science* **322**, 1819-1822 (2008).
22. K. Henzler *et al.*, Supersaturated calcium carbonate solutions are classical. *Sci. Adv.* **4**, eaao6283 (2018).
23. W. Sun, S. Jayaraman, W. Chen, K. A. Persson, G. Ceder, Nucleation of metastable aragonite $CaCO_3$ in seawater. *Proc. Nat. Acad. Sci. U.S.A.* **112**, 3199-3204 (2015).
24. D. B. Trushina, T. N. Borodina, S. Belyakov, M. N. Antipina, Calcium carbonate vaterite particles for drug delivery: Advances and challenges. *Mater. Today Adv.* **14**, (2022).
25. L. H. Fu, C. Qi, Y. R. Hu, C. G. Mei, M. G. Ma, Cellulose/vaterite nanocomposites: Sonochemical synthesis, characterization, and their application in protein adsorption. *Mater. Sci. Eng. C Mater. Biol. Appl.* **96**, 426-435 (2019).
26. R. E. Noskov *et al.*, Golden vaterite as a mesoscopic metamaterial for biophotonic applications. *Adv. Mater.* **33**, e2008484 (2021).
27. G. Steciuk, L. Palatinus, J. Rohlicek, S. Ouhenia, D. Chateigner, Stacking sequence variations in vaterite resolved by precession electron diffraction tomography using a unified superspace model. *Sci. Rep.* **9**, 9156 (2019).
28. A. G. Christy, A review of the structures of vaterite: the impossible, the possible, and the likely. *Cryst. Growth. Des.* **17**, 3567-3578 (2017).
29. R. Demichelis, P. Raiteri, J. D. Gale, R. Dovesi, The multiple structures of vaterite. *Cryst. Growth. Des.* **13**, 2247-2251 (2013).
30. E. Mugnaioli *et al.*, *Ab initio* structure determination of vaterite by automated electron diffraction. *Angew. Chem. Int. Ed.* **51**, 7041-7045 (2012).
31. L. Kabalah-Amitai *et al.*, Vaterite crystals contain two interspersed crystal structures. *Science* **340**, 454-457 (2013).
32. H. J. Meyer, Struktur und fehlordnung des vaterits. *Z. Kristallogr.* **128** 183-212 (1969).
33. A. Le Bail, S. Ouhenia, D. Chateigner, Microtwinning hypothesis for a more ordered vaterite model. *Powder Diffr.* **26**, 16-21 (2012).
34. R. Demichelis, P. Raiteri, J. D. Gale, R. Dovesi, A new structural model for disorder in vaterite from first-principles calculations. *CrystEngComm* **14**, 44-47 (2012).
35. L. Qiao, Q. L. Feng, Study on twin stacking faults in vaterite tablets of freshwater lacklustre pearls. *J. Cryst. Growth* **304**, 253-256 (2007).
36. S. J. Pennycook, D. E. Jesson, High-resolution incoherent imaging of crystals. *Phys. Rev. Lett.* **64**, 938-941 (1990).
37. Y. I. Svenskaya *et al.*, Key parameters for size- and shape-controlled synthesis of vaterite particles. *Cryst. Growth. Des.* **18**, 331-337 (2017).
38. X. H. Guo, S. H. Yu, G. B. Cai, Crystallization in a mixture of solvents by using a crystal modifier: morphology control in the synthesis of highly monodisperse $CaCO_3$ microspheres. *Angew. Chem. Int. Ed. Engl.* **45**, 3977-3981 (2006).
39. T. Schuler, W. Tremel, Versatile wet-chemical synthesis of non-agglomerated $CaCO_3$ vaterite nanoparticles. *Chem. Commun.* **47**, 5208-5210 (2011).
40. G. Behrens, L. T. Kuhn, R. Ubic, A. H. Heuer, Raman spectra of vateritic calcium carbonate. *Spectrosc. Lett.* **28**, 983-995 (1995).
41. U. Wehrmeister, A. L. Soldati, D. E. Jacob, T. Häger, W. Hofmeister, Raman spectroscopy of synthetic, geological and biological vaterite: a Raman spectroscopic study. *Journal of Raman Spectroscopy* **41**, 193-201 (2010).
42. S. Kamhi, On the structure of vaterite $CaCO_3$. *Acta Cryst.* **16**, 770-772 (1963).
43. L. Dupont, F. Portemer, t. late Michel Figlarz, Synthesis and study of a well crystallized $CaCO_3$ vaterite showing a new habitus. *J. Mater. Chem.* **7**, 797-800 (1997).

44. J. Wang, U. Becker, Structure and carbonate orientation of vaterite (CaCO$_3$). *Am. Mineral.* **94**, 380 (2009).
45. H. Wang, L. Zhang, J. Han, W. E, DeePMD-kit: A deep learning package for many-body potential energy representation and molecular dynamics. *Comput. Phys. Commun.* **228**, 178-184 (2018).
46. H. Niu, L. Bonati, P. M. Piaggi, M. Parrinello, *Ab initio* phase diagram and nucleation of gallium. *Nat. Commun.* **11**, 2654 (2020).
47. M. Yang, T. Karmakar, M. Parrinello, Liquid-liquid critical point in phosphorus. *Phys. Rev. Lett.* **127**, 080603 (2021).
48. W. Jia *et al.*, Pushing the limit of molecular dynamics with *ab initio* accuracy to 100 million atoms with machine learning. SC20: International conference for high performance computing, networking, storage and analysis. IEEE Press, 1-14 (2020).
49. A. Laio, M. Parrinello, Escaping free-energy minima. *Proc. Nat. Acad. Sci. U.S.A.* **99**, 12562-12566 (2002).
50. Y.-S. Ho, Review of second-order models for adsorption systems. *J. Hazard. Mater.* **136**, 681-689 (2006).
51. S. Radescu, I. Etxebarria, J. M. Perez-Mato, The Landau free energy of the three-dimensional Phi 4 model in wide temperature intervals. *J. Phys. Condens. Matter* **7**, 585 (1995).
52. M. Invernizzi, O. Valsson, M. Parrinello, Coarse graining from variationally enhanced sampling applied to the Ginzburg-Landau model. *Proc. Natl. Acad. Sci. U.S.A.* **114**, 3370-3374 (2017).
53. W. Jiang *et al.*, Chiral acidic amino acids induce chiral hierarchical structure in calcium carbonate. *Nat. Commun.* **8**, 15066 (2017).
54. J. Barthel, Dr. Probe: A software for high-resolution STEM image simulation. *Ultramicroscopy* **193**, 1-11 (2018).
55. G. Kresse, J. Hafner, *Ab initio* molecular dynamics for liquid metals. *Phys. Rev. B* **47**, 558-561 (1993).
56. G. Kresse, J. Furthmüller, Efficient iterative schemes for *ab initio* total-energy calculations using a plane-wave basis set. *Phys. Rev. B* **54**, 11169-11186 (1996).
57. P. E. Blöchl, Projector augmented-wave method. *Phys. Rev. B* **50**, 17953-17979 (1994).
58. G. Kresse, D. Joubert, From ultrasoft pseudopotentials to the projector augmented-wave method. *Phys. Rev. B* **59**, 1758-1775 (1999).
59. H. J. Monkhorst, J. D. Pack, Special points for Brillouin-zone integrations. *Phys. Rev. B* **13**, 5188-5192 (1976).
60. M. P. Allen, J. Tildesley D, *Computer simulation of liquids*. (Oxford University Press, 2017).
61. M. Parrinello, A. Rahman, Crystal structure and pair potentials: A molecular-dynamics study. *Phys. Rev. Lett.* **45**, 1196-1199 (1980).
62. M. Parrinello, A. Rahman, Polymorphic transitions in single crystals: A new molecular dynamics method. *J. Appl. Phys.* **52**, 7182-7190 (1981).
63. A. P. Thompson *et al.*, LAMMPS - a flexible simulation tool for particle-based materials modeling at the atomic, meso, and continuum scales. *Comput. Phys. Commun.* **271**, 108171 (2022).
64. G. Bussi, D. Donadio, M. Parrinello, Canonical sampling through velocity rescaling. **126**, 014101 (2007).
65. M. Bonomi *et al.*, Promoting transparency and reproducibility in enhanced molecular simulations. *Nat. Methods* **16**, 670-673 (2019).
66. A. Barducci, G. Bussi, M. Parrinello, Well-tempered metadynamics: A smoothly converging and tunable free-energy method. *Phys. Rev. Lett.* **100**, 020603 (2008).
67. A. Stukowski, Visualization and analysis of atomistic simulation data with OVITO–the Open Visualization Tool. *Modell. Simul. Mater. Sci. Eng.* **18**, 015012 (2010).
68. K. Momma, F. Izumi, VESTA 3 for three-dimensional visualization of crystal, volumetric and morphology data. *J. Appl. Crystallogr.* **44**, 1272-1276 (2011).

69. G. Bussi, A. Laio, Using metadynamics to explore complex free-energy landscapes. *Nat. Rev. Phys.* **2**, 200-212 (2020).

**Acknowledgements:** The authors are grateful to Prof. Qingling Feng at Tsinghua University for kindly providing the bulk pearl samples and Dr. Chi Zhang in Ocean University of China for many helpful discussions on vaterite structures.

**Funding:** X.S. acknowledges the National Natural Science Foundation of China (No. 51901065), the Nature Science Foundation of Hebei Province (No. E2020201023), and the Advanced Talents Incubation Program of Hebei University (No. 521000981164) for financial support. H.N. was supported by the National Science Fund for Excellent Young Scientist Fund Program (Overseas) of China, the Science and Technology Activities fund for Overseas Researchers of Shanxi Province, and the Research Fund of the State Key Laboratory of Solidification Proceeding (NPU) of China (No. 2020-QZ-03). This work made use of the EPIC facility of Northwestern University's NU*ANCE* Center, which has received support from the SHyNE Resource (NSF ECCS-2025633), the IIN, and Northwestern's MRSEC program (NSF DMR-1720139).

**Contributions**: X.S. and X.H. performed the microscopy experiments and data analysis. J.H., M.C. and H.N. performed the theoretical calculations. X.H. and H.N. conceived and coordinated the entire project. All authors contributed to the discussion and writing of the manuscript.

**Competing interests:** The authors declare that they have no competing interests.

**Data and materials availability:** All data needed to evaluate the conclusions of this work are presented in the paper and the Supplementary Materials. Additional data related to this paper may be requested from the authors.

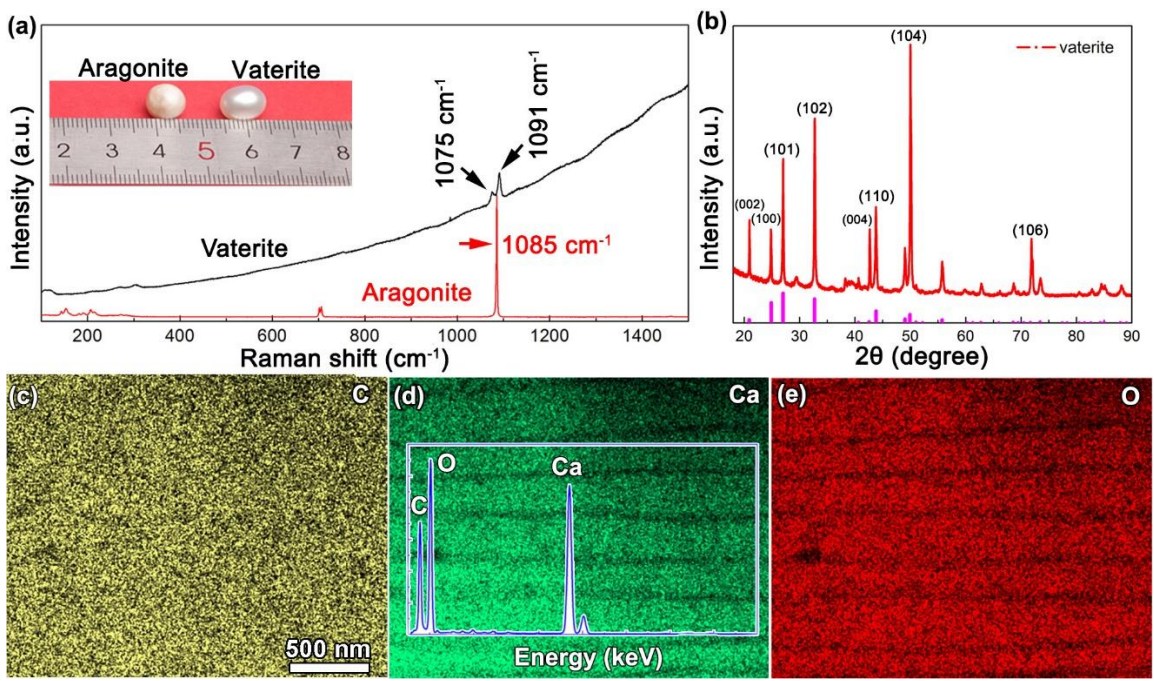

**Extended Data Fig. 1. Microstructure and chemical features of aragonite.** (a) Raman spectra of the pearls with aragonite and vaterite structures. The inset microscopic image shows the general appearance of the lackluster and normal pearls. (b) XRD patterns of the freshwater lackluster pearl. Bottom vertical pink lines show the peak positions of vaterite structure from PDF-33-0628. (c-e) C, Ca and O maps of vaterite, corresponding to the region shown in Fig. 1b. The inset in (d) is the EDS spectrum of vaterite.

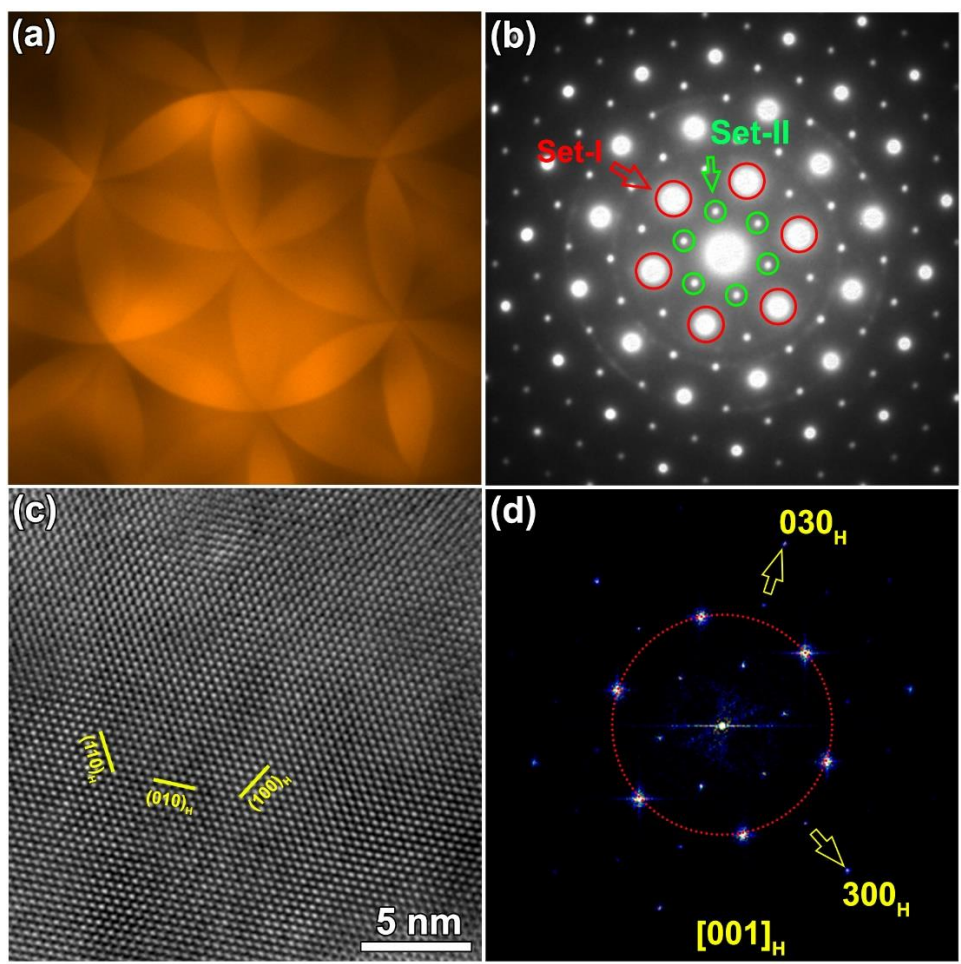

**Extended Data Fig. 2. Pseudo-hexagonal symmetry within vaterite.** (a) Convergent beam electron diffraction patterns of vaterite along the same direction axis as Fig. 2b. (b) Duplicate EDPs of Fig. 2b. Red circles indicate some of the Set-I patterns with higher intensity. Green circles indicate some of the Set-II patterns with relatively lower intensity. (c) Atomic resolution TEM images of vaterite along the $[103]_m$ direction. (d) Digital fast Fourier transform (FFT) patterns of the HRTEM image in (c).

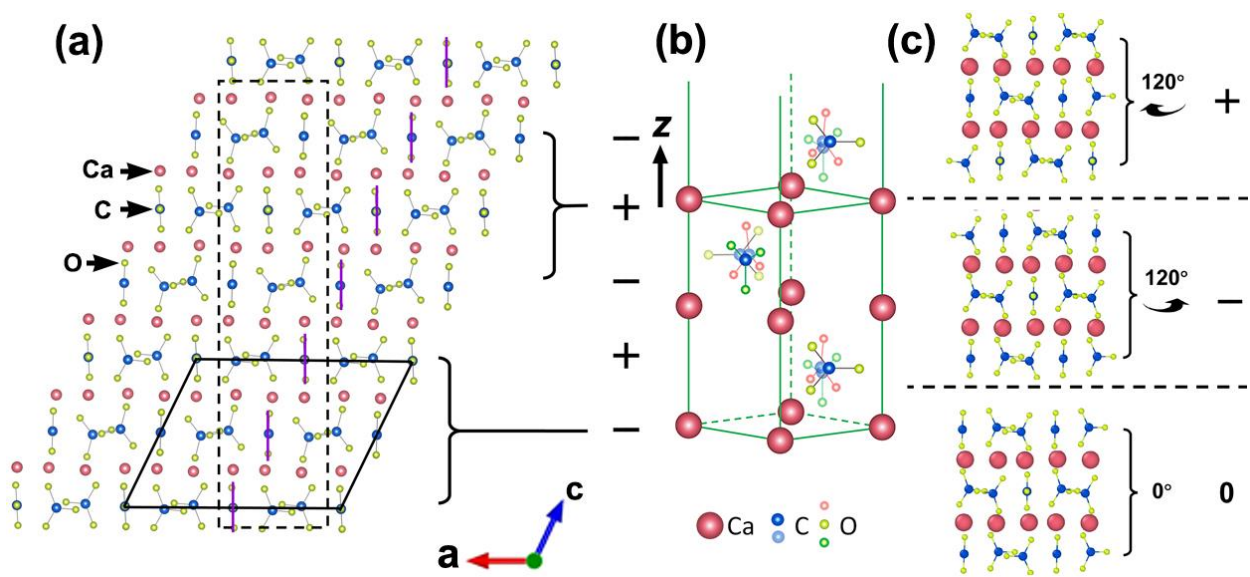

**Extended Data Fig. 3. Stacking features of vaterite.** (a) Structural projection of the monoclinic *C2* phase along the [010] direction. The shot purple lines indicate the stacking features of carbonate layers. The unit cell is denoted by the solid black lines, while the dashed box encloses a hexagonal lattice description of the *C2* structure. (b) Stereo schematic showing the general arrangements of calcium atoms and carbonates in vaterite. One of the C-O bonds should point towards the edge of the trigonal prism, resulting in three possible orientations of carbonates that occupy half of the $Ca_6$ trigonal prismatic interstices. (c) The possible stacking sequences composed of three adjacent layers, which can be marked as "+", "−", and "0", respectively.

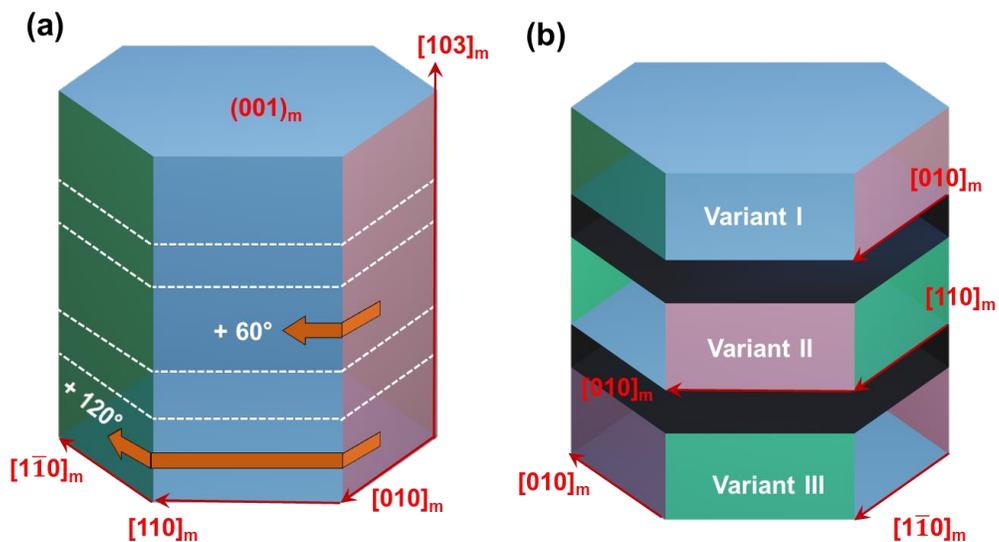

**Extended Data Fig. 4. Polytypic features within vaterite.** Schematic illustration showing the formation of the orientation variant. (a) Pristine structure with uniform orientations. (b) Polytypic structure with variants. Variants II and III can be obtained by continuous 60° rotation of the carbonate layers from variant I along the $[103]_m$ direction. The dark area in (b) indicates the disordered stacking between different variants.

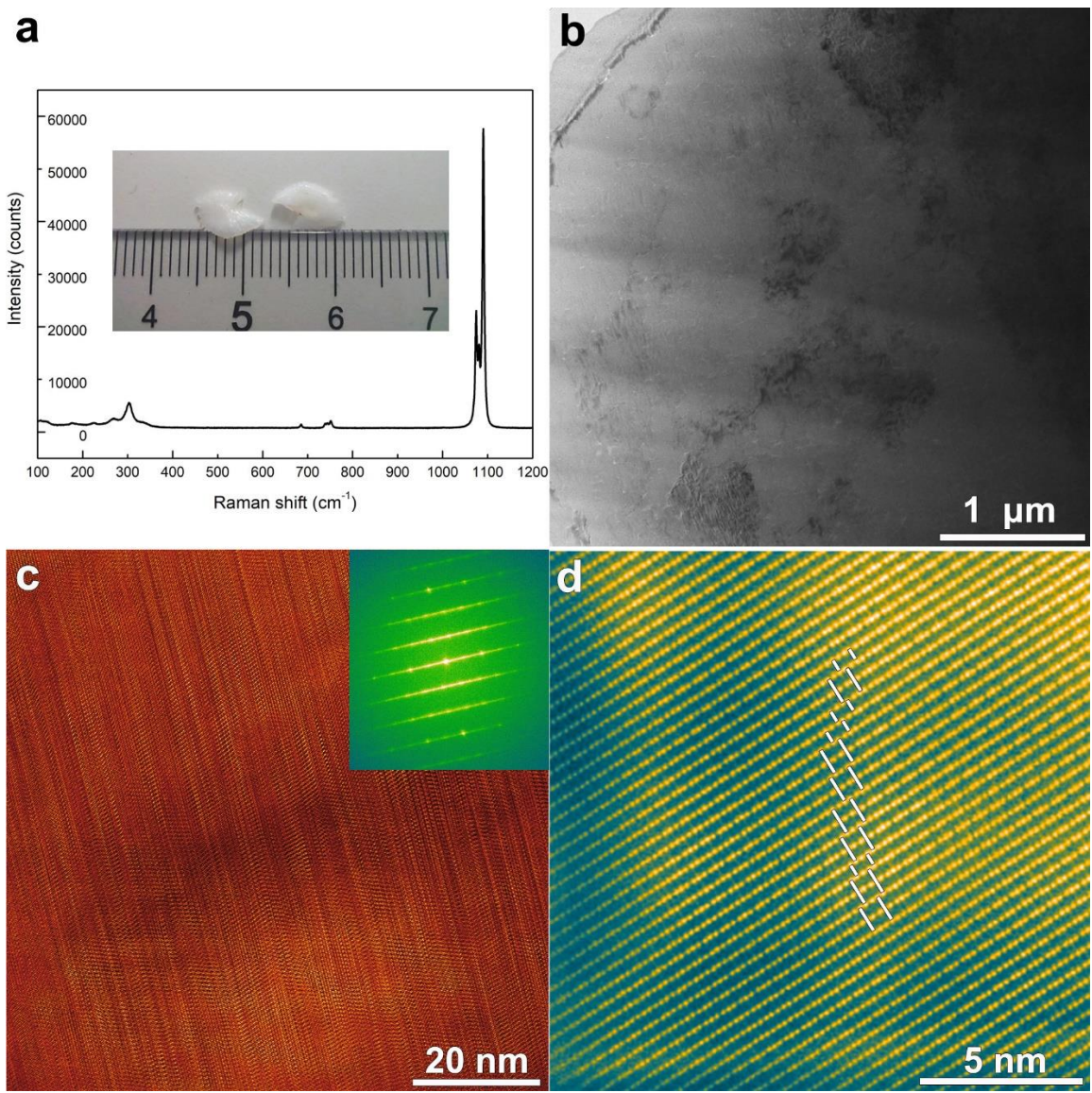

**Extended Data Fig. 5. Microstructural features for the asteriscus pairs of carp having vaterite structure.** (a) Raman spectrum. Inset is the optical image. (b) Low magnification TEM image. (c) HRTEM image and inset corresponding FFT patterns along [010]$_m$ zone axis again showing the representative faulted feature in vaterite. (d) Atomic resolution HAADF image along the [010]$_m$ zone axis showing the polytypic features of vaterite. The lines indicate the stacking feature of the Ca-C-O chains along the [001]$_m$ direction.

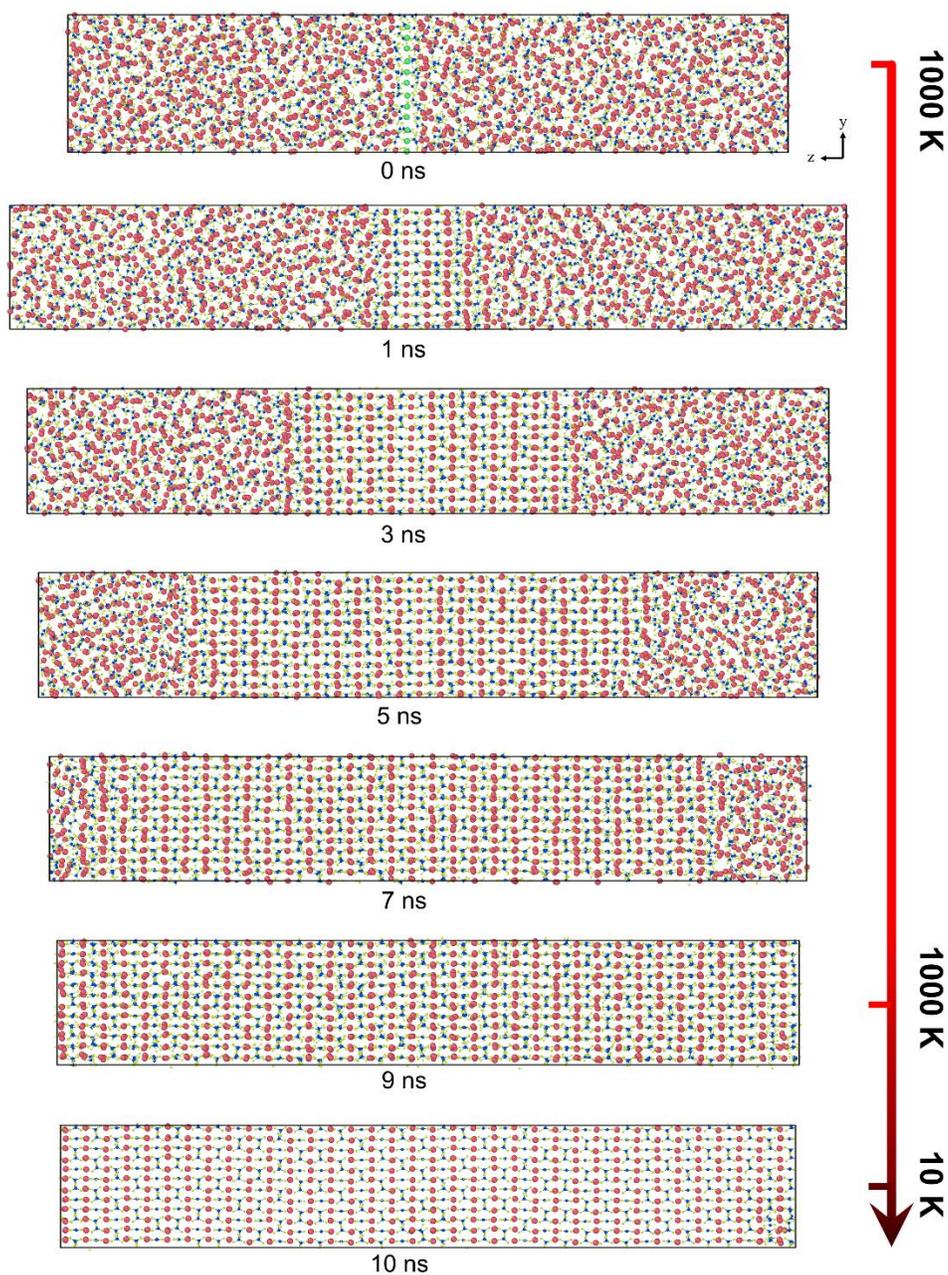

**Extended Data Fig. 6. Crystallization simulation of vaterite growth from liquid.** Snapshots of molecular dynamics simulations of vaterite growing along the *z*-axis direction during the first 9 *ns* at 1000 *K* after fixing a layer of calcium atoms and quenching to 10 *K* during the last 1 *ns*. The simulation cell contains 6480 atoms (1296 $CaCO_3$ units). The green atoms in the snapshot of 0 *ns* indicate the fixed layer of calcium atoms. The pink, blue, and yellow–green spheres represent calcium, carbon, and oxygen atoms, respectively.

# Unlocking the Structural Mystery of Vaterite CaCO$_3$

Xingyuan San[1†], Junwei Hu[2†], Mingyi Chen[2], Haiyang Niu[2*], Paul J. M. Smeets[3], Jie Deng[4], Kunmo Koo[3], Roberto dos Reis[3], Vinayak P. Dravid[3*], Xiaobing Hu[3*]

[1]Hebei Key Lab of Optic-electronic Information and Materials, The College of Physics Science and Technology, Hebei University, Baoding 071002, China

[2]State Key Laboratory of Solidification Processing, International Center for Materials Discovery, School of Materials Science and Engineering, Northwestern Polytechnical University, Xi'an 710072, China

[3]Department of Materials Science and Engineering, The NU*ANCE* Center, Northwestern University, Evanston, IL 60208, USA

[4]Department of Geosciences, Princeton University, Princeton, NJ 08544, USA

[†]These authors contributed equally to this work.

[*]Corresponding authors. Email: xbhu@northwestern.edu (X.H.); haiyang.niu@nwpu.edu.cn (H.N.); v-dravid@northwestern.edu (V.P.D.)

**This PDF file includes:**

Supplementary Text (Note I to VIII)
Figs. S1 to S17
Tables S1 to S2
References (1 to 23)

**Supplementary Text**

**Note I. Controversy on Structure of Vaterite**

Regarding the structure of vaterite, there are many controversies in the literature (Table S1). Approximately one hundred years ago, researchers found that the structure of vaterite was distinct from that of calcite and aragonite based on *X*-ray diffraction (XRD) analyses (*1*). Then, the cell with hexagonal symmetry (*a=4.11 Å* and *c=8.513 Å*) was derived in 1925 by Olshausen (*2*). Based on further XRD analysis, Meyer provided the first crystallographic description *(3)* and proposed an orthorhombic cell model with the space group of *Pnma* and lattice parameters of *a=4.13 Å*, *b=7.15 Å, c=8.48 Å*. Several years later, Kamhi revealed that the vaterite structure has a hexagonal symmetry with space group *P6₃/mmc* and lattice parameters of *a=4.13 Å*, *c=8.49 Å* using the same XRD technique (*4*). Within Kamhi's hexagonal cell, there is only one unique carbonate group inside the cell, and each atom of the $CO_3^{2-}$ ion is disordered with a partial occupancy of 1/3, which accounts for considerable disorder. In 1969, Meyer provided a superstructure based on Kamhi's structure, which has the same space group but different carbonate site symmetry (*5*). He believed that the possible stacking disorder of single layers of trigonal symmetry was responsible for diffuse streaks and satellite reflections in the diffraction pattern. Since then, researchers have gradually realized that conventional lab XRD techniques or even advanced synchrotron beam-based diffraction techniques are very limited in solving this structure. Recently, researchers have attempted to address this structural mystery by means of other advanced complementary methods such as theoretical calculations and diffraction tomography, which have resulted in a number of new structure models based on a micro-twinning hypothesis (*Ama2*) (*6*), DFT calculations (*P3₂21*) (*7*), automated diffraction tomography (*C2/c* and *P1̄*) (*8*), molecular dynamics simulations (*P6₅22*) (*9*), high-resolution transmission electron microscopy (HRTEM) data of a biogenic vaterite sample (*P6₃mmc* of Kamhi-cell in 1963 and unknown cell) (*10*) and precession electron diffraction tomography (*C12/c1*) (*11*). Furthermore, spectroscopic techniques such as nuclear magnetic resonance (NMR) and Raman spectroscopy are also used to verify the crystal structure of vaterite (*12-15*). NMR spectra show that the *P3₂21* (or *P3₁21*) model and the monoclinic *C2* model provide the best simultaneous agreement between the experimental and calculated NMR data (*12*). The Raman spectrum shows that there are at least three structurally independent carbonate groups in the unit cell of vaterite (*13*). More complicated, researchers gradually accepted that vaterite cannot be described based on a single lattice and that there are likely plenty of planar faults including micro-twining and stacking disorder (*6, 11, 16*).

**Table S1** Structure information of vaterite CaCO$_3$

| Space Group | Lattice Parameters | Methods |
|---|---|---|
| ***Pbnm*, *Pnma*** | a=4.13 Å, b=7.15 Å, c=8.48 Å<br>α=β=γ=90° | XRD (*3*), theory (*17*) |
| ***P6$_3$22*** | a=b=7.14 Å, c=8.52 Å<br>α=β=90°, γ=120° | XRD (*18*) |
| ***P6$_3$/mmc*** | a=b=4.13 Å, c=8.49 Å<br>α=β=90°, γ= 120° | XRD (*4, 6*) |
| ***P6$_3$/mmc*** | a=b=7.169 Å, c=16.98 Å<br>α=β= 90°, γ=120° | XRD (*5, 19*) |
| ***P6$_5$22, P3$_5$21, P6$_5$*** | a=b=7.29 Å, c=25.30 Å<br>α=β= 90°, γ=120° | XRD (*20*), theory (*7, 9, 21*) |
| ***Ama2*** | a=8.47 Å, b=7.16 Å, c=4.13 Å<br>α=β=γ=90° | Theory (*6*) |
| ***P2$_1$2$_1$2$_1$*** | a = 4.37 Å, b = 6.58 Å, c = 8.43 Å<br>α=β=γ=90° | Theory (*7*) |
| ***C2, Cc, C2/c*** | a=12.17 Å, b=7.12 Å, c = 9.47Å<br>α=γ=90°, β=118.94° | ED (*8*), theory (*22*) |
| ***C1, C$\bar{1}$*** | a = 12.36 Å, b = 7.11 Å, c = 25.74 Å<br>α = 90.43°, β = 99.88°, γ = 90.29° | ED (*8*), theory (*22*) |

**Note II. Different marking methods for vaterite**

In addition to various unit cells proposed to explain the crystal structure of vaterite, two marking methods were developed to describe both ordered and disordered models with any stacking sequence according to the characteristics of each carbonate layer (*22, 23*). The planes of carbonates with half of the specific occupation of triangular prisms made up by the calcium atoms are labelled with the letters A, B, and C, while each letter indicates an orientation of carbonates. Another half of the occupation with carbonate planes is labelled with A′, B′ or C′ in line with the different orientations of carbonates. Fig. S2a presents the projections of a stacking model onto the *xz*-plane of the six possible permutations of carbonates while odd-number layers and even-number layers are denoted by different colors. The carbonate layers with the same color are labelled with the same type of letters, e.g., (A, B, C) or (A′, B′, C′), which correspond to half of the specific occupation as shown in Fig. S2b. However, this marking method has its own drawbacks. By rotating the vaterite along the *z*-axis with an integer multiple of 60°, a structure with calcium atoms and carbonates like the structure before rotation is obtained since the calcium atoms exhibit a pseudohexagonal lattice. However, the letters marked for each layer will not retain their sequences as the structure rotates. In other words, each direction requires a certain set of letters to label the sequence of carbonates. For instance, the labels of the vaterite with the *C2* space group in the three directions are shown in Fig. S3a. We find that the corresponding labels of the projections obtained in different directions are not the same, even though they share the same stacking sequence.

In Christy's work (*23*), the relation of orientations between two layers of carbonate separated by one another is used to label the stacking sequence. The marking method is shown in Extended Data Fig. 3. In this way, the two layers of carbonates separated by one another apart from +120°, -120° or 0° can be marked as "+", "−", and "0" while the middle layer is indifferent. For instance, the *C2* structure and $P3_221$ structure can be marked by '+−+−+−' and '++++++', respectively, while reflection in a vertical mirror plane will interconvert + and – operations. This method can describe any stacking sequence well with the projections of different directions. Fig. S3b shows the labels of the vaterite with the *C2* space group in three directions, while their labels are identical.

**Note III. Experimental Indistinguishability of Three Basic Monoclinic Structures**

**Table S2.** Extinction rule for the space groups of *C2*, *Cc* and C*2/c*.

|      | (hkl)   | (h0l)   | (0kl)  | (hk0)   | (h00)  | (0k0)  | (00l)  |
|------|---------|---------|--------|---------|--------|--------|--------|
| C2   | h+k=2n  | h=2n    | k=2n   | h+k=2n  | h=2n   | k=2n   |        |
| Cc   | h+k=2n  | h, l=2n | k=2n   | h+k=2n  | h=2n   | k=2n   | l=2n   |
| C2/c | h+k=2n  | h, l=2n | k=2n   | h+k=2n  | h=2n   | k=2n   | l=2n   |

In Demichelis's work (*22*), a basic monoclinic structure with three possible space groups (No.5, *C2*; No.9, *Cc*; No.15, *C2/c*) have been proposed based on theoretical calculations. Although the proposed basic monoclinic structure has three possible space groups, their lattice parameters are similar to each other with *a=12.3* Å, *b=7.13* Å, *c=9.4* Å, *β*=115.48°. Regarding the space groups of *C2*, *Cc* and *C2/c*, it is impossible to distinguish them based on the electron diffraction patterns (EDPs). The extinction rule of space groups *C2*, *Cc* and *C2/c* in EDPs are listed in Table S2, where *n* equals positive integer. The space groups *Cc* and *C2/c* share the same extinction rule. Thus, the diffraction patterns should have the same distributions. That is why we cannot distinguish them based on EDPs. Meanwhile, considering the double diffraction resulting from the dynamic effect, we cannot distinguish *C2* from *Cc* or *C2/c*, although there are some slight differences in their extinction rules. The simulated EDPs of the vaterite structure with the above three space groups along some low index directions are shown in Fig. S4 and Fig. S5. It is found that the distribution of diffraction patterns with above three space groups along [010], [103], [001], and [101] are the same. Only along the [100] direction, there are slight differences (see Fig. S4a, 4c, 4e). In the simulated EDPs, (00$l$) with $l$ equal to an odd number does not exist in Fig. S4a-d but occurs in Fig. S4e, 4f. However, due to the unavoidable dynamic effects when electron beams interact with the samples, these patterns should always exist. Thus, we cannot distinguish them based on the distributions of the diffraction patterns. Indeed, there may be slight differences in the intensities of the reflection patterns for different space groups. However, it is very challengeable to quantify the intensities of diffractions in conventional electron diffraction since the intensities depend on many factors such as the beam intensity, sample thickness and orientation deviations of local areas. Therefore, it is very challenging to distinguish the above three possible vaterite structures based on only the experimental EDPs.

**Note IV. Inherent Relationships between the Monoclinic Lattice and Hexagonal Lattice**

The disordered Meyer structure with a hexagonal lattice (*P6$_3$/mmc*; $a=b=7.169$ Å, $c=16.98$ Å) has been extensively discussed (*5, 19*). Here, based on elaborate crystallographic considerations, we uncover the inherent relationships between the abovementioned monoclinic lattice and Meyer's hexagonal lattice. Since we cannot distinguish the space groups among *C2*, *Cc*, and *C2/c* experimentally, here we simply labeled them as a monoclinic lattice. As shown in Fig. S6, along the [103]$_m$ direction, where the subscript *m* represents the monoclinic lattice, the projection of the monoclinic lattice has a pseudohexagonal feature. The lengths of the [010]$_m$, [110]$_m$ and [1$\bar{1}$0]$_m$ directions are approximately the same. Meanwhile, since the Meyer's hexagonal structure has 6-fold symmetry along the [001]$_H$ direction, where the subscript H represents the hexagonal lattice, it is reasonable to deduce that there is similarity between the lattices of Meyer's hexagonal structure and monoclinic structures. Interestingly, it is found that the lengths of the [010]$_m$, [100]$_m$, and [206]$_m$ directions are almost the same as those of the [010]$_H$, [210]$_H$, and [003]$_H$ directions, as indicated in Fig. S6. Thus, the inherent orientation relationships between the hexagonal Meyer lattice and monoclinic lattice can be described using the following matrix transformation.

$$\begin{pmatrix} u \\ v \\ w \end{pmatrix}_H = \begin{pmatrix} 2 & 0 & -\frac{2}{3} \\ 1 & 1 & -\frac{1}{3} \\ 0 & 0 & \frac{1}{2} \end{pmatrix} \begin{pmatrix} u \\ v \\ w \end{pmatrix}_m$$

## Note V. Possible Orientational Variants within the Monoclinic Lattice

Considering the existence of the pseudohexagonal feature within the monoclinic lattice along the $[103]_m$ direction, as shown in Fig. S6, it is reasonable to deduce that there may be three possible orientation variants within the monoclinic structure, which is formed due to the 60° (or 120°) rotation along the $[103]_m$ zone-axis. As revealed in Fig. S6, because of continuous 60° clockwise rotation, the $[010]_m$ direction will change to the $[110]_m$ direction and then to the $[1\bar{1}0]_m$ direction. Meanwhile, the $[200]_m$ direction will change to the $[1\bar{3}0]_m$ direction and then to the $[\bar{1}\bar{3}0]_m$ direction. Thus, the directions within three variants (V1, V2, V3) can be correlated using the following matrix transformation.

$$\begin{pmatrix} u \\ v \\ w \end{pmatrix}_{V2} = \begin{pmatrix} \frac{1}{2} & \frac{1}{2} & \frac{1}{6} \\ -\frac{3}{2} & \frac{1}{2} & \frac{1}{2} \\ 0 & 0 & 1 \end{pmatrix} \begin{pmatrix} u \\ v \\ w \end{pmatrix}_{V1}$$

$$\begin{pmatrix} u \\ v \\ w \end{pmatrix}_{V3} = \begin{pmatrix} \frac{1}{2} & -\frac{1}{2} & \frac{1}{6} \\ \frac{3}{2} & \frac{1}{2} & -\frac{1}{2} \\ 0 & 0 & 1 \end{pmatrix} \begin{pmatrix} u \\ v \\ w \end{pmatrix}_{V1}$$

## Note VI. Index of Serial Electron Diffraction Patterns

Intuitively, all the EDPs shown in Fig. 2a-f have a clear feature that the intensity of the reflections is not uniform. There are clearly 2 sets of patterns in each figure. As indicated in Fig. S15a which is a local magnification of Fig. 2a, one set of patterns has a very high intensity (Set-I) indicated by blue circles, and the other set (Set-II) of patterns has a relatively weak intensity indicated by red circles. The intensity of Set-I patterns is >10 times larger than that of the Set-II patterns. Here, we first only consider the relative positions of the reflections but ignore the intensity difference. Simulations of the EDPs of Meyer's hexagonal structure and the monoclinic structure along some major zone-axes are shown in Fig. S7-S9. Detailed analysis demonstrates that the EDPs shown in Fig. 2a can be tentatively indexed based on the hexagonal Meyer structure. However, the remaining EDPs in Fig. 2 can only be indexed based on the monoclinic structure. The reason why Fig. 2a cannot be indexed based on the monoclinic structure is the existence of Set-II weak patterns. For example, Set-I patterns such as $\{030\}_H$, $\{300\}_H$ and $\{330\}_H$ in Fig. 2a can be indexed as the $[103]_m$ zone-axis based on the monoclinic lattice (see Fig. S8a). However, Set-II patterns such as $\{100\}_H$, $\{200\}_H$, $\{110\}_H$ and $\{220\}_H$ should not occur along the $[103]_m$ zone-axis.

The relative distributions of the EDPs shown in Fig. 2b, 2d and 2f are the same except for a 60° rotation along the viewing direction. All of them can be indexed based on the monoclinic structure and correspond to the $[001]_m$ zone-axis. Considering the experimental tilting angles shown in Fig. S2, it is found that the intersection angle between positions B (D) and D (F), which correspond to the EDPs of Fig. 2b (2d) and 2d (2f) respectively, is 60°. The relative distribution of EDPs shown in Fig. 2c and 2e are the same as well, and the intersection angle between position C (corresponding to Fig. 2c) and E (corresponding to Fig. 2e) is 60° as well. Thus, it is deduced that there are likely three orientation variants within the vaterite structure. These three orientation variants are generated due to a 60° rotation along the $[103]_m$ zone-axis, which agrees well with the above crystallographic considerations in Note V. The orientation relationship among variants I, II and III can be described using the matrix listed in Note V. When variant I is tilted to the $[001]_m^{V1}$ direction, variants II and III will be orientated along the $[136]_m^{V2}$ and $[1\bar{3}6]_m^{V3}$ directions respectively. The simulated EDPs along the $[001]_m^{V1}$, $[136]_m^{V2}$ and $[1\bar{3}6]_m^{V3}$ directions are shown in Fig. S7a, 7c, and 7e, respectively. By overlapping them, the simulated patterns reproduce the experimental results shown in Fig. 2b, 2d and 2f. Set-I patterns with higher intensity are shared by variants I, II and III. However, Set-II patterns only occur in variant I along the $[001]_m^{V1}$ direction. Thus, Set-I patterns can have a higher intensity and Set-II patterns have a relatively lower intensity. Similarly, when variant I is tilted along the $[101]_m^{V1}$ direction, variants II and III will be orientated to the $[2\bar{3}3]_m^{V2}$ and $[233]_m^{V3}$ directions respectively. Overlapping their simulated EDPs (Fig. S7b, 7d, 7f) will reproduce the experimental results shown in Fig. 2c and 2e perfectly.

Now, let us move on to the EDPs shown in Fig. 2a again, which can be tentatively labelled as [001]$_H$ for simplicity. Since the hexagonal Meyer structure and monoclinic structure are closely related, the [001]$_H$ direction in the hexagonal Meyer structure is equivalent to the [103]$_m$ direction in the monoclinic structure (Fig. S6). Since variants I, II, III are formed due to a 60° rotation along the [103]$_m$ zone-axis, EDPs along the [103]$_m$ direction for variants I-III should be the same (see Fig. S8a). Overlapping the [103]$_m$ EDPs can only generate the Set-I patterns. However, the occurrence of the Set-II patterns in the hexagonal Meyer structure is due to the disordered arrangement of carbonate ($CO_3^{2-}$). Along the [103]$_m$ direction, after continual 60° rotation, three types of carbonates seem to be arranged in a "disordered" way. Thus, this will introduce additional Set-II patterns, which are confirmed by the simulated EDPs shown in Fig. 15b based on our theoretically grown vaterite with polytypic structural features.

**Note VII. Structural and Energy Characteristics of Carbonates with Different Stackings**

Here, we will clarify how the stacking orders of layers with different carbonate orientations along the z-axis affect the stability of vaterite. As stated in the main text, the carbonates in one layer have three different possible orientations. If only looking into the stacking of two contiguous layers, any stacking order is equivalent due to rotational symmetry, as shown in Fig. S10. Nevertheless, if we consider the stacking orders of two carbonate layers separated by one another, as introduced in Extended Data Fig.3b, the situation is quite different. As shown in Fig. S10b and 10c, when the two layers of carbonates have different orientations, corresponding to the stacking order of "+" or "−", the carbonates and calcium atoms are arranged like a honeycomb on the projection of the *xy*-plane. On the other hand, as shown in Fig. S10d, when the two layers of carbonates have the same orientation, corresponding to the stacking order of "0", their projections on the *xy*-plane are apparently coincident.

The above discussion has illustrated that the scenario of the stacking order "0" is entirely different from that of "+" or "−". Such difference can be further elucidated from the perspective of energy. We run MD simulations with the initial configurations corresponding to different scenarios using the DNN model at 300 *K*, and their potential energies as a function of time are shown in Fig. S14. One can find that if the stacking sequences only contain "+" and/or "-", the average energies are almost the same (within 2.5 *meV/f.u.*, less than the error margin of the DNN model). In contrast, the energy of the stacking sequence containing only "0" is approximately 41 *meV/f.u.* higher than the former ones. The result demonstrates that the sequence order "0" is energetically unfavorable.

## Note VIII. High-Symmetry Structure and Low-Symmetry Structure Identification and Collective Variable

In previous work, Demichelis *et al.* (7, 22) proposed several low-symmetry structures (LS) with greater stability based on the density functional theory (DFT) calculations, while the earlier proposed high-symmetry structures (HS) correspond to the transition state. The difference between HS and LS is the tilt angle of the carbonates, which also leads to a slight shift in the positions of calcium atoms.

According to the different tilt angles of carbonates, we therefore define the difference between the coordinates of the two overlapping oxygen atoms of one layer in the *z*-axis direction as $\Delta z_1$ as shown in Fig. S17. Since the tilting feature of carbonates occurs in two layers as a period in the *z*-axis direction, the coordinate difference in the *z*-axis direction of two overlapping oxygen atoms of adjacent layers is defined as $\Delta z_2$ according to the same definition. We define $\Delta z_1$-$\Delta z_2$ as the collective variable (CV) to describe the symmetry of our system. This definition has the advantage that when the CV is zero, the structure corresponds to HS. When the CV is not zero, the structure corresponds to LS. However, using only a few carbonates to define the symmetry of the system will inevitably lead to large errors. Therefore, we use this method to redefine the order parameter by using the coordinate information of all overlapping carbonates and oxygen atoms in the system such as

$$Q = \frac{2}{mn} \sum_{i=1}^{n} \sum_{k=1}^{\frac{m}{2}} (\Delta z_{2k-1,i} - \Delta z_{2k,i}),$$

where *m* is the number of layers of carbonates in the *z*-axis of the system, and *n* is the number of pairs of carbonates with two overlapping oxygen atoms in one layer. In this way, two-thirds of the carbonates in the system are used for the CV calculations.

**Supplementary Figures**

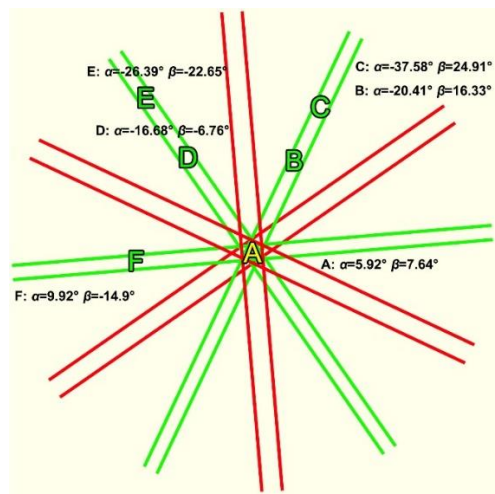

**Fig. S1.** Original experimental tilting data for the serial EDPs shown in Fig. 2.

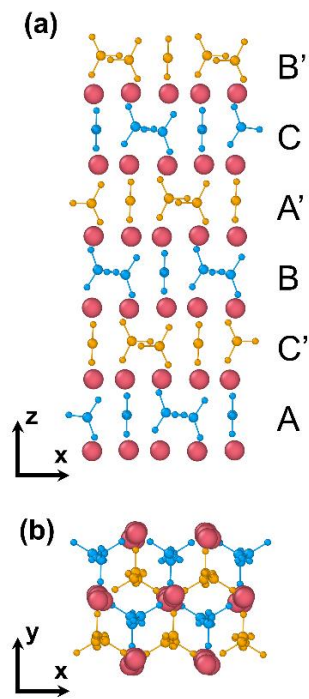

**Fig. S2.** Structural projections of a stacking model on the (a) *xz*-plane and (b) *xy*-plane. In this model, all six possible permutations of carbonates are labelled with different letters. The carbonates in the odd layers and even layers are colored blue and brown, respectively. The carbonates with the same color are approximately ±120° or 0° apart from each other in the *xy*-plane.

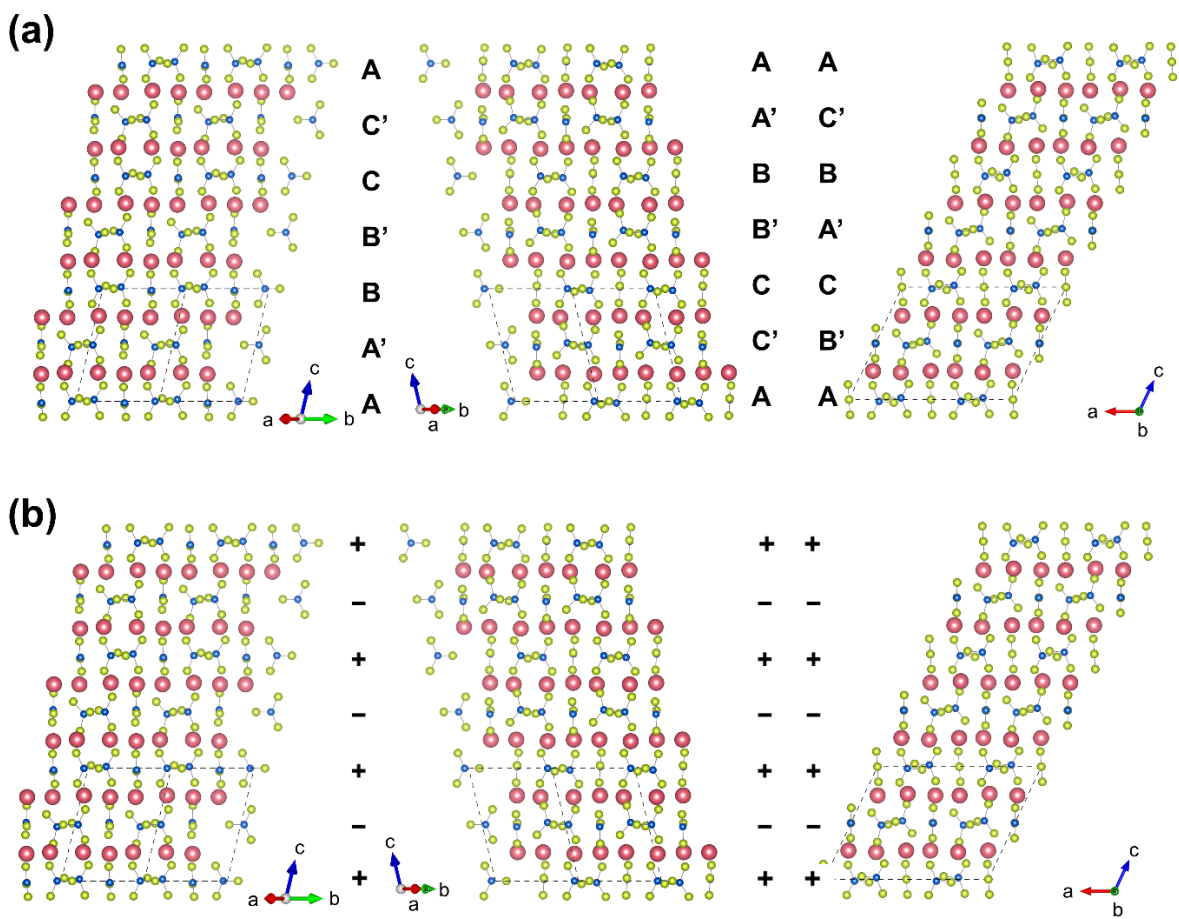

**Fig. S3.** The projections of vaterite structure with *C2* space group along the [0$\bar{1}$0], [$\bar{1}\bar{1}$0] and [1$\bar{1}$0] directions. The stacking order of carbonates was labelled according to different marking methods, while the method used in (a) comes from Demichelis' work (*22*) and (b) comes from Christy's work (*23*).

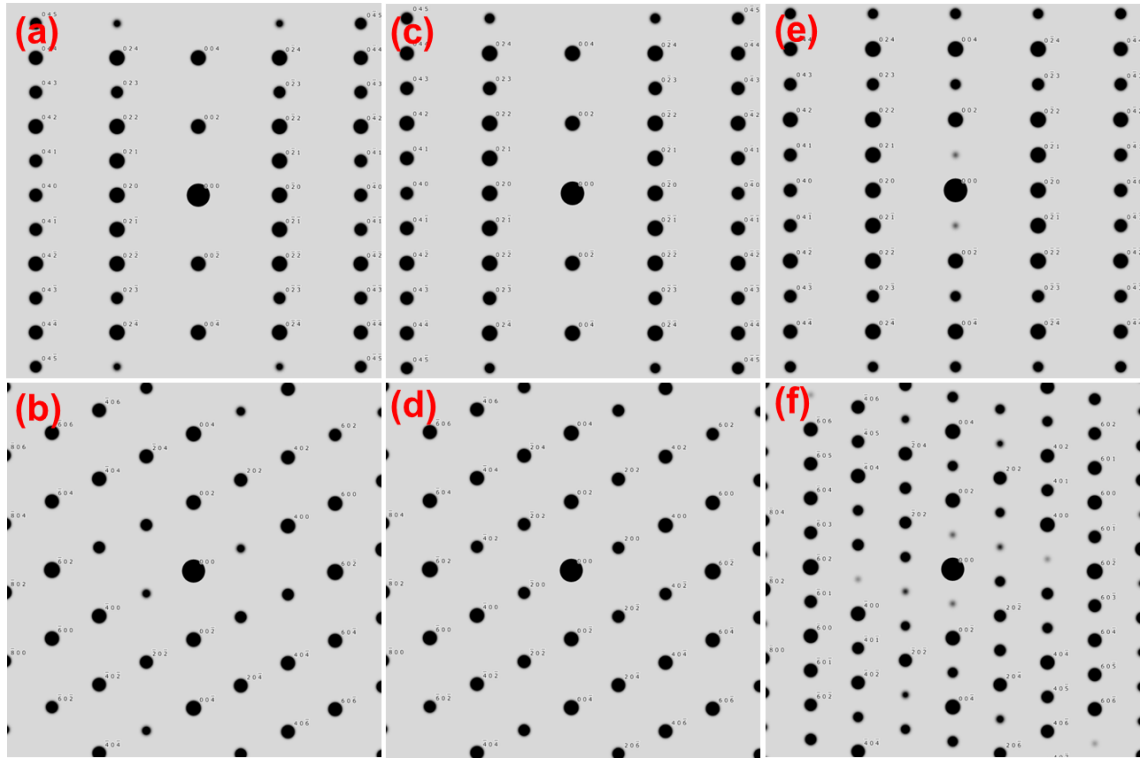

**Fig. S4.** Simulated EDPs of the vaterite CaCO$_3$ with (a, b) the space group of *C2/c* along the [100], [010] directions; (c, d) the space group of *Cc* along the [100], [010] directions; (e,f) the space group of *C2* along the [100], [010] directions.

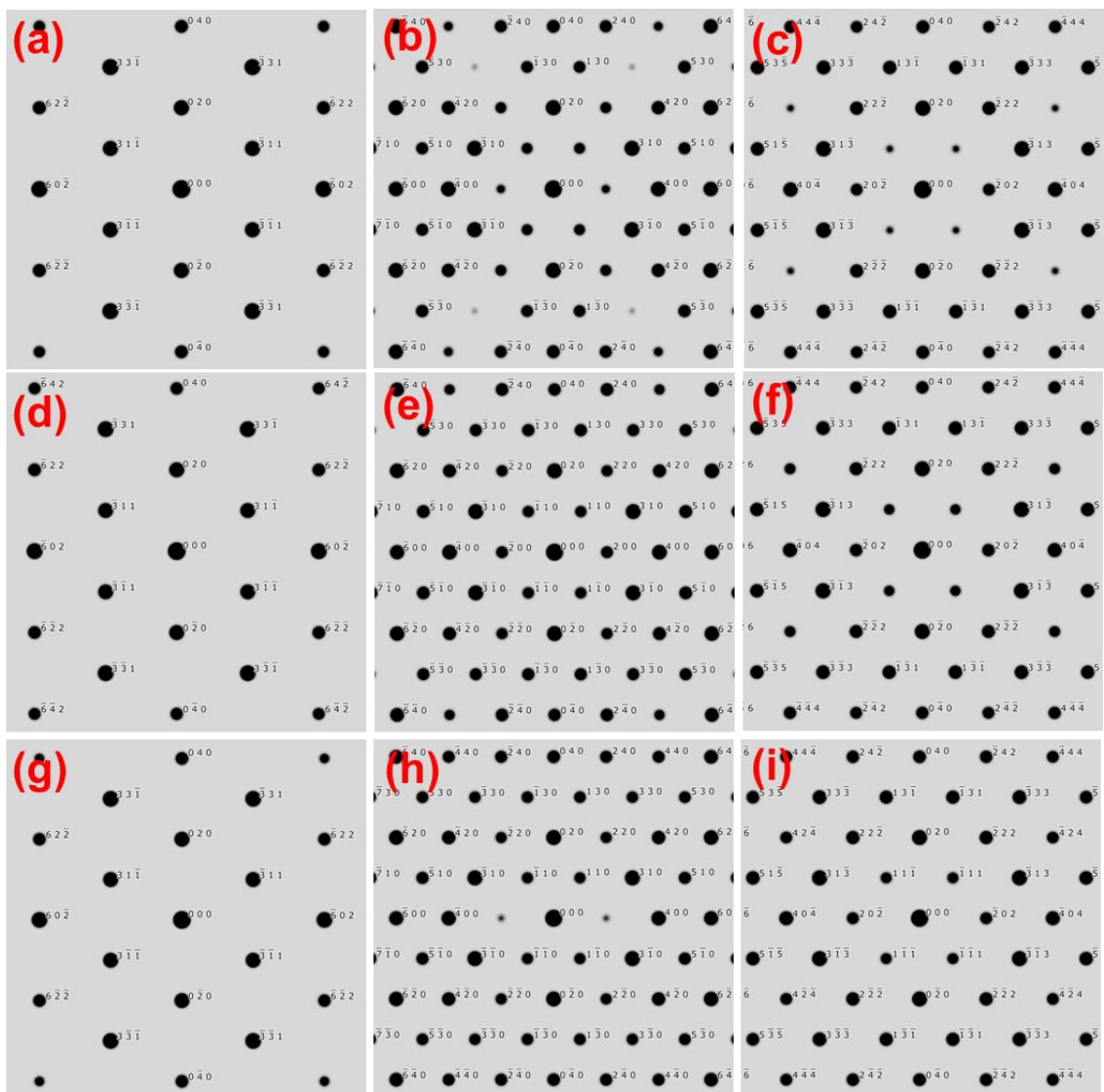

**Fig. S5.** Simulated EDPs of the vaterite CaCO$_3$ with (a-c) the space group of *C2/c* along the [103], [001] and [101] directions; (d-f) the space group of *Cc* along the [103], [001] and [101] directions; (g-i) the space group of *C2* along the [103], [001] and [101] directions.

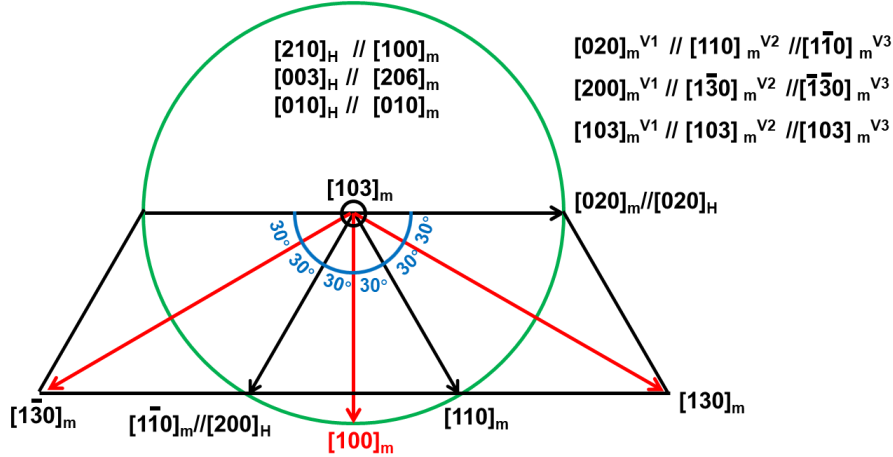

**Fig. S6.** Schematic showing the orientation relationship between the disordered Meyer structure with the space group of *P*6$_3$/*mmc* and the monoclinic structure with the space group of *C2/c* or *Cc* or *C2*. The subscripts m and H indicate monoclinic latter and hexagonal lattices, respectively. The superscripts V1, V2 and V3 indicate three possible orientational variants within the monoclinic lattice.

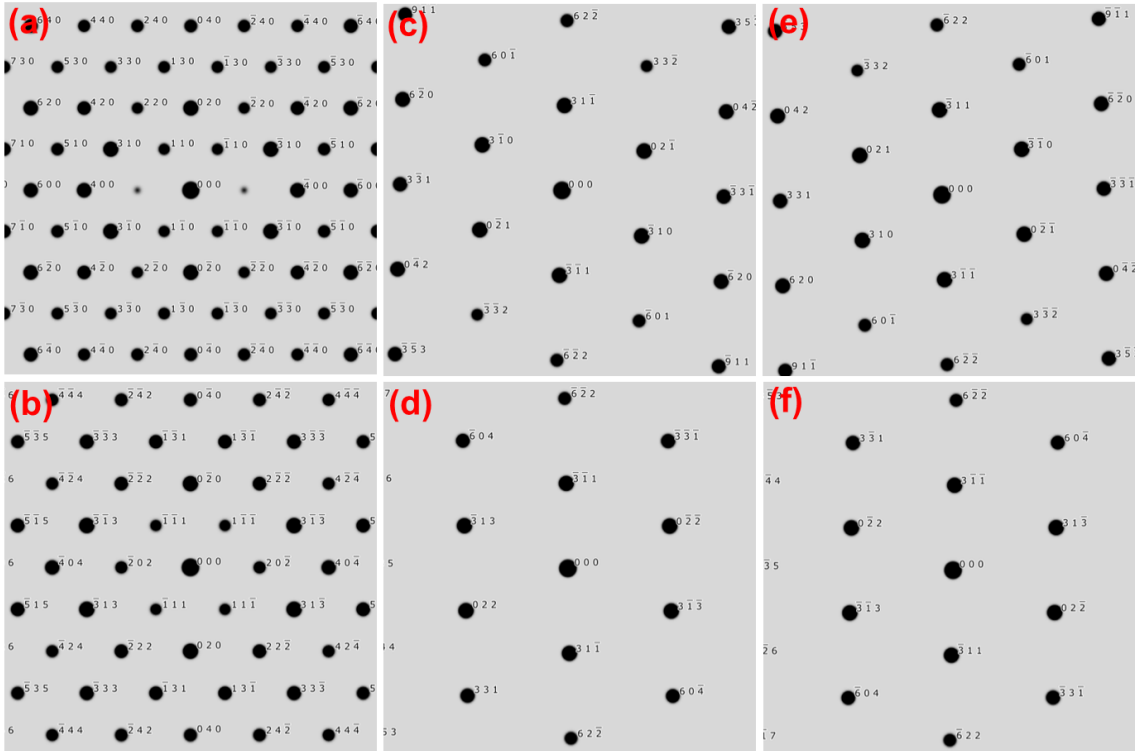

**Fig. S7.** Simulated EDPs of (a, b) the ordered *C2*-V1 along the $[001]_m^{V1}$, $[101]_m^{V1}$ directions, (c, d) the ordered *C2*-V2 along the $[136]_m^{V2}$, $[2\bar{3}3]_m^{V2}$ directions, and (e, f) the ordered *C2*-V3 along the $[1\bar{3}6]_m^{V3}$, $[233]_m^{V3}$ directions.

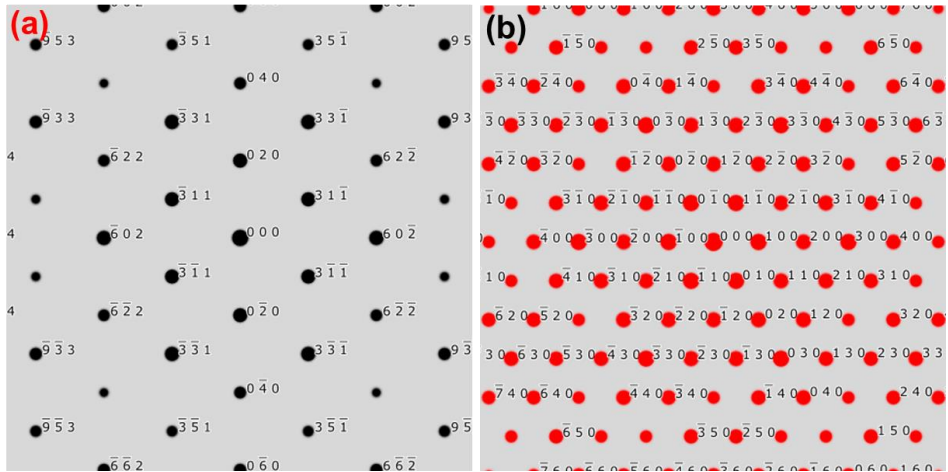

**Fig. S8.** Simulated EDPs of (a) the ordered *C2* structure along the $[103]_m$ direction and (b) the disordered hexagonal Meyer structure along the $[001]_H$ direction.

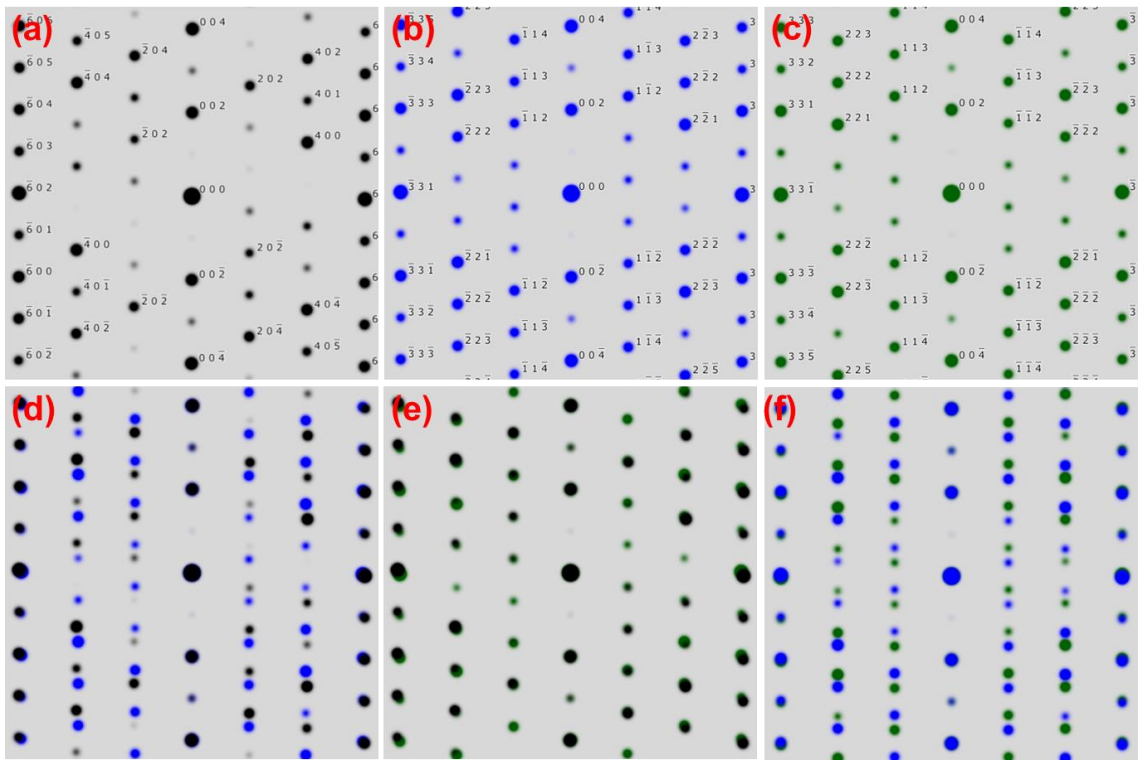

**Fig. S9.** Simulated EDPs of the ordered *C2* vaterite along (a) $[010]_m^{V1}$, (b) $[110]_m^{V2}$, (c) $[1\bar{1}0]_m^{V3}$ and composite EDPs of (d) $[010]_m^{V1}$ // $[110]_m^{V2}$, (e) $[010]_m^{V1}$ // $[1\bar{1}0]_m^{V3}$, (f) $[110]_m^{V2}$// $[1\bar{1}0]_m^{V3}$ direction.

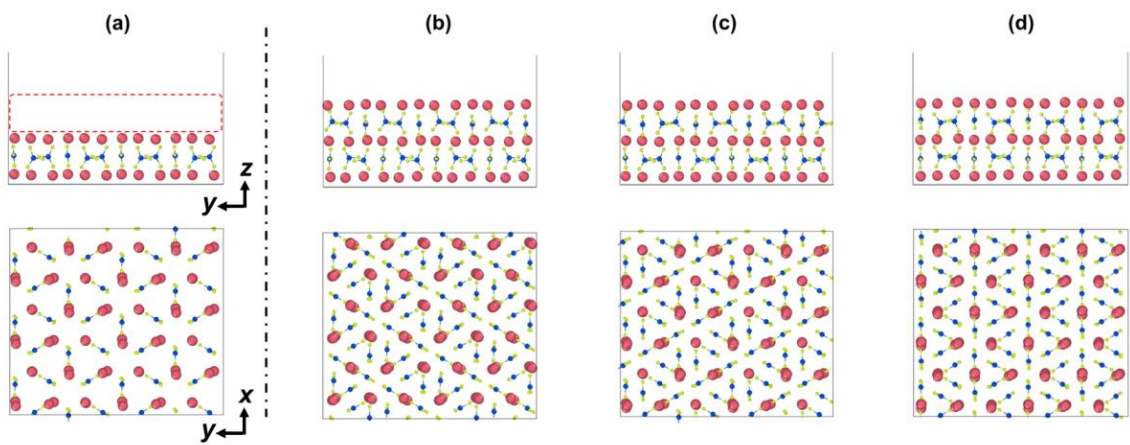

**Fig. S10.** (a) Structural projections of only one-layer carbonates in the box on the *yz*-plane (top) and *xy*-plane (bottom). (b-d) The projections of three possible structures after placing the second layer of carbonates on the next layer along the *z*-direction on the *yz*-plane (top) and *xy*-plane (bottom). The pink, blue, and yellow–green spheres represent calcium atoms, carbon atoms and oxygen atoms, respectively.

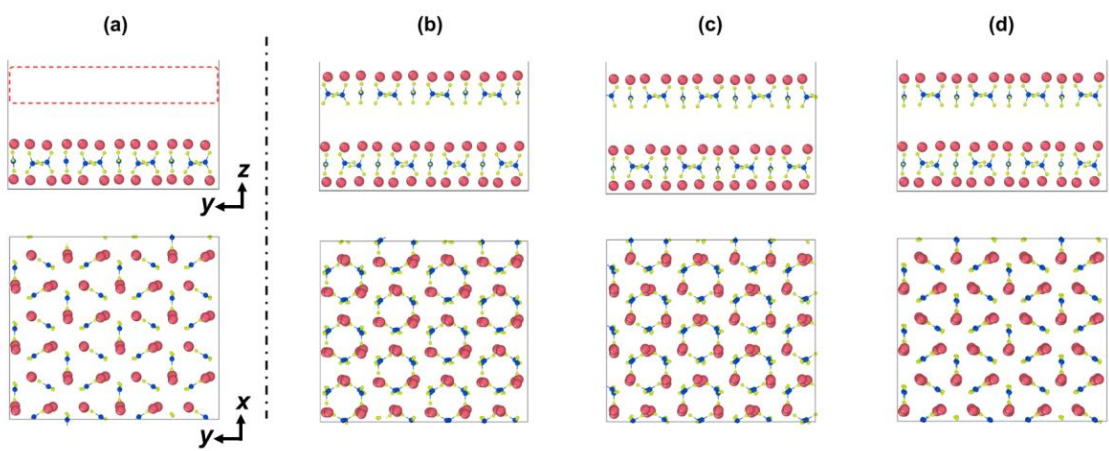

**Fig. S11.** (a) Structural projections of only one-layer carbonates in the box on the *yz*-plane (top) and *xy*-plane (bottom). (b-d) The projections of three possible structures after placing another layer of carbonates on the third layer of carbonates along the *z*-direction on the *yz*-plane (top) and *xy*-plane (bottom). The pink, blue, and yellow–green spheres represent calcium atoms, carbon atoms and oxygen atoms, respectively.

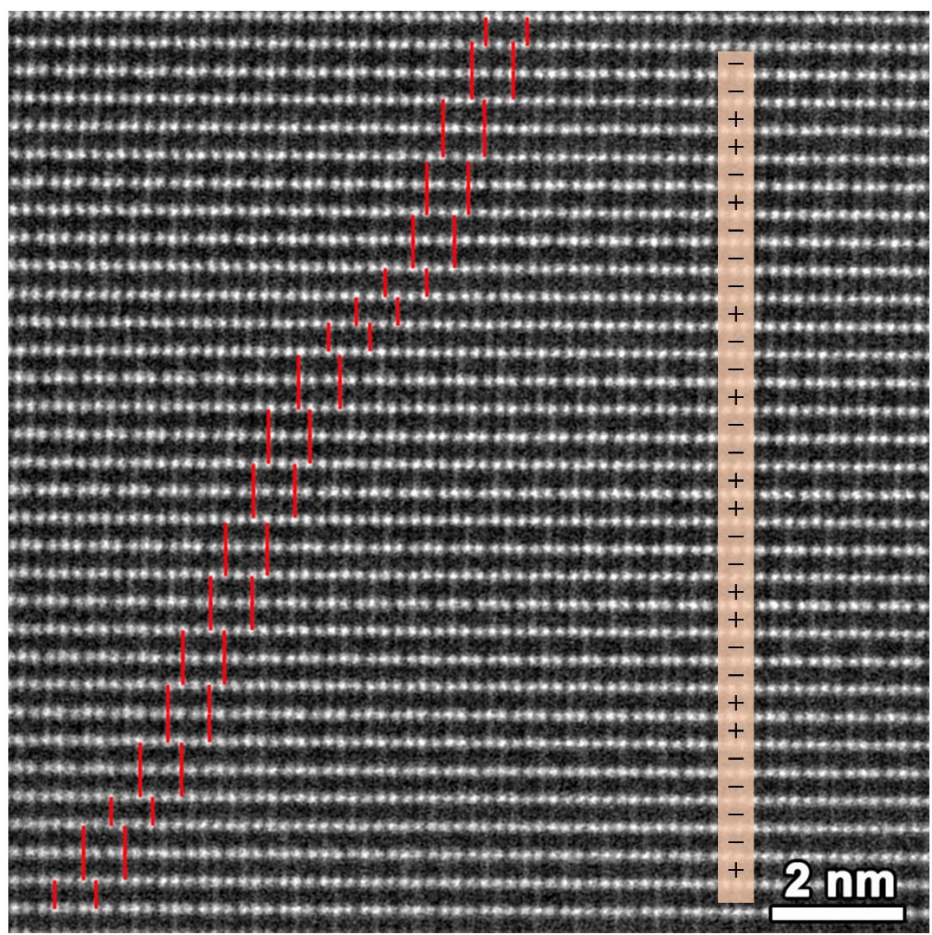

**Fig. S12.** Atomic resolution HAADF image along the $[010]_m$ direction showing the polytypic features within vaterite along the $[001]_m$ direction. The red vertical lines indicate the stacking feature of the Ca-C-O chains along the $[001]_m$ direction. The inserted symbols indicate the stacking sequences of the carbonate layers introduced in Extended Data Fig. 3.

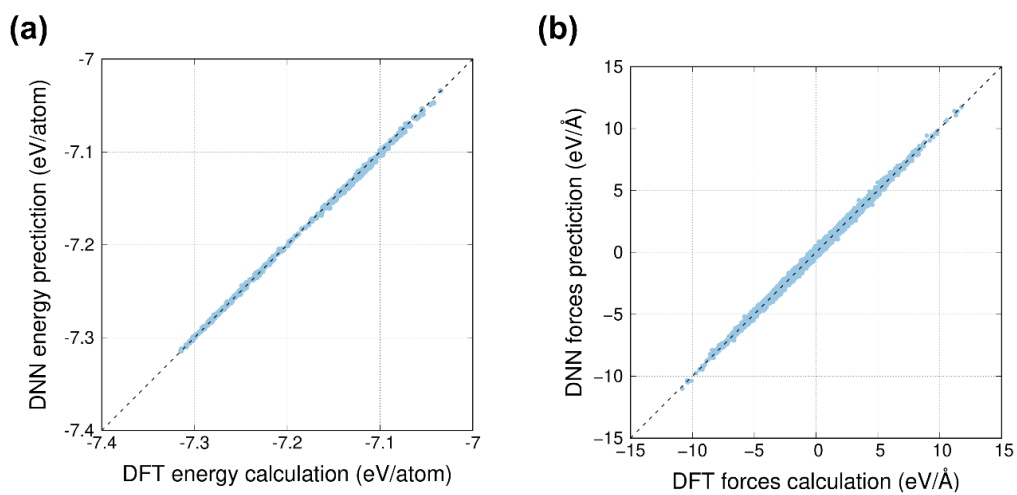

**Fig. S13.** Comparison of atomic energy (a) and forces (b) on the test set between the DFT calculation and deep neural network (DNN) model prediction. The configurations in the test set were extracted from MD trajectories using the DNN model under corresponding thermodynamic conditions.

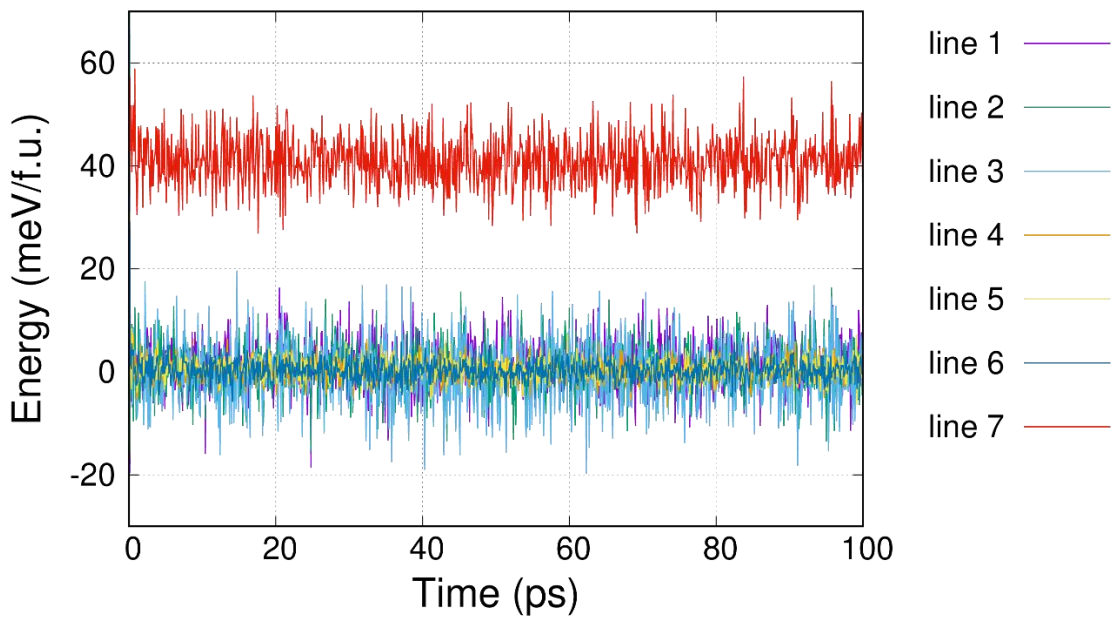

**Fig. S14.** The energy at 300 $K$ for structures with different carbonate orientations as a function of time. The line 7 represents the structure with the stacking sequence of "000000" (1080 atoms), while others represent the structures with the stacking sequence containing only "+" and/or "−". (line 1: "++++++" (1080 atoms), line 2: "+−+−+−" (1080atoms), line 3: "++−−" (720 atoms), line 4, line 5 and line 6 represent the 36-layers structures (6480 atoms) obtained by MD simulations with disordered stacking sequences.)

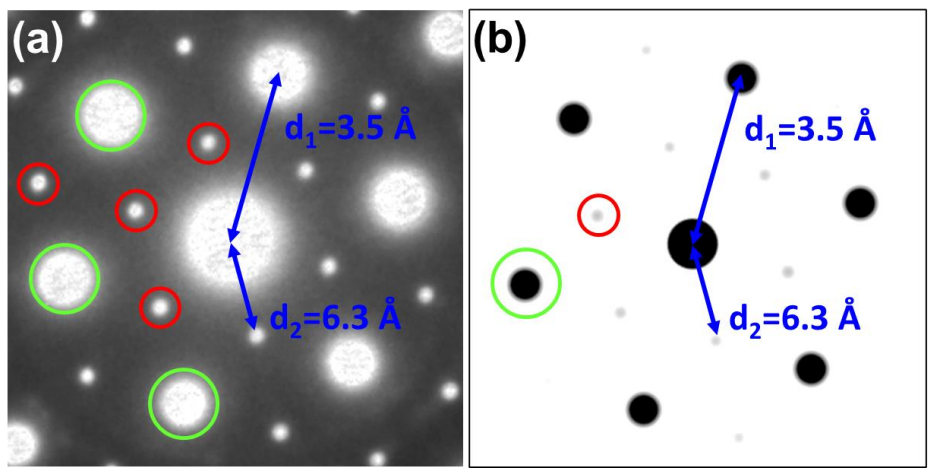

**Fig. S15.** (a) Local magnification of Fig. 2a EDPs. (b) Simulated EDPs of the theoretically grown polytypic structure along the stacking direction.

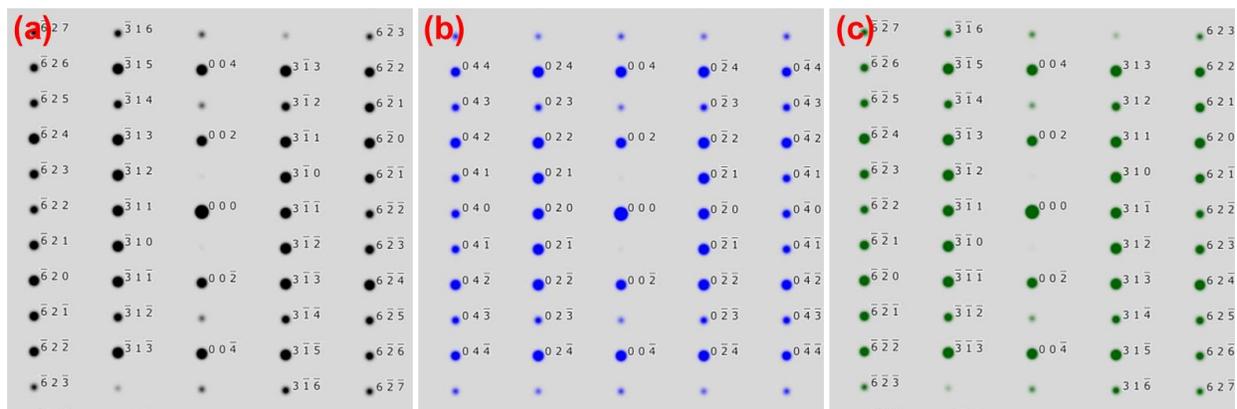

**Fig. S16.** Simulated EDPs of the ordered *C2* vaterite along the (a) $[130]_m^{V1}$, (b) $[100]_m^{V2}$ and (c) $[1\bar{3}0]_m^{V3}$ directions.

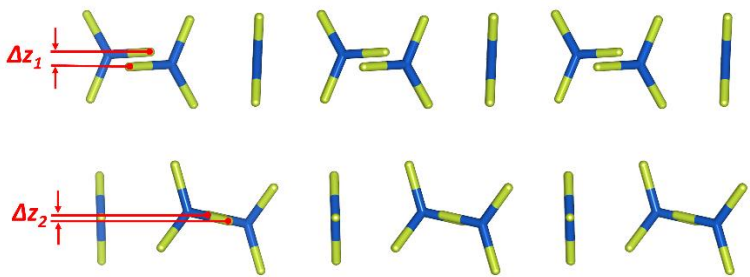

**Fig. S17**. Schematic diagram showing how collective variable are defined. The carbonates are the fragment in the *C2* structure with low symmetry.

# References

1. R. E. Gibson, R. W. G. Wyckoff, H. E. Merwin, Vaterite and μ-calcium carbonate. *Am. J. Sci.* **s5-10**, 325-333 (1925).
2. S. Olshausen, Vaterite hexagonal cell determination by X-ray powder pattern. *Z. Kristallogr.* **61**, 463-464 (1925).
3. H. J. Meyer, Uber vaterit und seine struktur. *Angew. Chem., Int. Ed.* **71**, 678-678 (1959).
4. S. Kamhi, On the structure of vaterite $CaCO_3$. *Acta Cryst.* **16**, 770-772 (1963).
5. H. J. Meyer, Struktur und fehlordnung des vaterits. *Z. Kristallogr.* **128** 183-212 (1969).
6. A. Le Bail, S. Ouhenia, D. Chateigner, Microtwinning hypothesis for a more ordered vaterite model. *Powder Diffr.* **26**, 16-21 (2012).
7. R. Demichelis, P. Raiteri, J. D. Gale, R. Dovesi, A new structural model for disorder in vaterite from first-principles calculations. *CrystEngComm* **14**, 44-47 (2012).
8. E. Mugnaioli *et al.*, *Ab initio* structure determination of vaterite by automated electron diffraction. *Angew. Chem. Int. Ed.* **51**, 7041-7045 (2012).
9. J. Wang, U. Becker, Structure and carbonate orientation of vaterite ($CaCO_3$). *Am. Mineral.* **94**, 380 (2009).
10. L. Kabalah-Amitai *et al.*, Vaterite crystals contain two interspersed crystal structures. *Science* **340**, 454-457 (2013).
11. G. Steciuk, L. Palatinus, J. Rohlicek, S. Ouhenia, D. Chateigner, Stacking sequence variations in vaterite resolved by precession electron diffraction tomography using a unified superspace model. *Sci. Rep.* **9**, 9156 (2019).
12. K. M. Burgess, D. L. Bryce, On the crystal structure of the vaterite polymorph of $CaCO_3$: a calcium-43 solid-state NMR and computational assessment. *Solid State Nucl. Magn. Reson.* **65**, 75-83 (2015).
13. U. Wehrmeister, A. L. Soldati, D. E. Jacob, T. Häger, W. Hofmeister, Raman spectroscopy of synthetic, geological and biological vaterite: a Raman spectroscopic study. *Journal of Raman Spectroscopy* **41**, 193-201 (2010).
14. C. Gabrielli, R. Jaouhari, S. Joiret, G. Maurin, *In situ* Raman spectroscopy applied to electrochemical scaling. Determination of the structure of vaterite. *J. Raman Spectrosc.* **31**, 497-501 (2000).
15. D. L. Bryce, E. B. Bultz, D. Aebi, Calcium-43 chemical shift tensors as probes of calcium binding environments. insight into the structure of the vaterite $CaCO_3$ polymorph by 43Ca solid-state NMR spectroscopy. *J. Am. Chem. Soc.* **130**, 9282-9292 (2008).
16. L. Qiao, Q. L. Feng, Study on twin stacking faults in vaterite tablets of freshwater lacklustre pearls. *J. Cryst. Growth* **304**, 253-256 (2007).
17. S. K. Medeiros, E. L. Albuquerque, F. F. Maia, E. W. S. Caetano, V. N. Freire, First-principles calculations of structural, electronic, and optical absorption properties of $CaCO_3$ Vaterite. *Chem. Phys. Lett.* **435**, 59-64 (2007).
18. J. D. C. McConnell, Vaterite from Ballycraigy, Larne, Northern Ireland. *Miner. Mag. J. Miner. Soc.* **32**, 535-544 (1960).
19. L. Dupont, F. Portemer, t. late Michel Figlarz, Synthesis and study of a well crystallized $CaCO_3$ vaterite showing a new habitus. *J. Mater. Chem.* **7**, 797-800 (1997).
20. B. C. Chakoumakos, B. M. Pracheil, R. P. Koenigs, R. M. Bruch, M. Feygenson, Empirically testing vaterite structural models using neutron diffraction and thermal analysis. *Sci. Rep.* **6**, 36799 (2016).

21. J. Wang *et al.*, Carbonate orientational order and superlattice structure in vaterite. *J. Cryst. Growth* **407**, 78-86 (2014).
22. R. Demichelis, P. Raiteri, J. D. Gale, R. Dovesi, The multiple structures of vaterite. *Cryst. Growth. Des.* **13**, 2247-2251 (2013).
23. A. G. Christy, A review of the structures of vaterite: the impossible, the possible, and the likely. *Cryst. Growth. Des.* **17**, 3567-3578 (2017).